\documentclass[aps,pre,reprint]{revtex4-2}

\usepackage[utf8]{inputenc}
\usepackage[T1]{fontenc}
\usepackage{amsmath, amssymb,subfigure}
\usepackage{graphicx}
\usepackage[colorlinks,allcolors=blue]{hyperref}

\newcommand{\davg}{\langle d \rangle}
\newcommand{\savg}{\langle\langle \sigma \rangle\rangle}

\begin{document}
\title{Structural crossovers of quasi-one-dimensional patchy hard superellipses}

\author{Sakineh Mizani}
\email{sakineh.mizani@mnf.uni-tuebingen.de}
\affiliation{Institute for Applied Physics, University of T\"ubingen, Auf der Morgenstelle 10, 72076 T\"ubingen, Germany}

\author{P\'eter Gurin}
\email{gurin.peter@mk.uni-pannon.hu}
\affiliation{Physics Department, Centre for Natural Sciences,
University of Pannonia, P.O. Box 158, Veszpr\'em H-8201, Hungary}

\author{Martin Oettel}
\email{martin.oettel@uni-tuebingen.de}
\affiliation{Institute for Applied Physics, University of T\"ubingen, Auf der Morgenstelle 10, 72076 T\"ubingen, Germany}

\author{Szabolcs Varga}
\email{varga.szabolcs@mk.uni-pannon.hu}
\affiliation{Physics Department, Centre for Natural Sciences,
University of Pannonia, P.O. Box 158, Veszpr\'em H-8201, Hungary}

\begin{abstract}
We study a quasi-one-dimensional associating fluid composed
of hard superellipses carrying two patches interacting
through a directional Kern--Frenkel potential.
Using the Transfer Operator Method, we show that the
selective patch--patch association promotes horizontal
alignment and chain formation at low-to-intermediate
densities, whereas hard-core interaction favours vertical
alignment without bonds at high densities.
The competition between these two mechanisms drives a
structural crossover upon compression from a horizontally
aligned bonded chain structure to a completely unbonded,
vertically aligned structure.
While patchy ellipses undergo a tilted-to-vertical
realignment, patchy rectangle-like superellipses exhibit a
horizontal-to-vertical change.
These structural changes manifest as a plateau in the
equation of state.
To capture these properties, we generalise Wertheim's
first-order thermodynamic perturbation theory by introducing
an orientation-dependent fraction of sites not in a bond.
When combined with the Parsons--Lee hard-body theory, the
orientationally resolved perturbation theory provides
quantitatively reliable results for the structural
properties and phase behaviour. Therefore, the generalised Wertheim theory together with Parsons-Lee theory can be suitable in higher dimensions, too. 
\end{abstract}
 
\maketitle

\section{Introduction}

Self-assembly is a fundamental process in which elementary
building blocks spontaneously organise into complex,
well-defined structures through local interparticle
interactions, enabling the bottom-up formation of functional
materials that exhibit unique collective
properties~\cite{Glotzer2007,Zhang2014}. A central advance has been the recognition that particles
decorated with a small number of discrete, chemically
distinct surface sites---so-called patches---exhibit
directional and selective interactions that drive assembly
into programmed geometries, in direct analogy with atomic
valence in molecular systems~\cite{Wang2012,Li2020}.
Therefore, understanding how microscopic directional interactions give
rise to macroscopic order and emergent material properties
remains a central challenge in soft-matter physics, addressed
through experiments, statistical-mechanical theory, and
simulation~\cite{Bianchi2011,DeMichele2012,Sacanna2013}, and increasingly
through machine-learning and inverse-design
approaches~\cite{Dijkstra2021}.
 
In three dimensions, patchy and associating particles exhibit
a wide variety of self-assembled structures determined by the
number, geometry, and selectivity of their bonding sites.
Bifunctional particles form polymer chains and may
undergo an isotropic--nematic liquid-crystal transition as a
function of concentration and bonding
strength~\cite{DeMichele2012,Lu2004,Lu2006,Nguyen2018}. Moreover, patchy particles with reduced valence or anisotropic
geometry can assemble into micelles and vesicles in analogy with
surfactant systems~\cite{Kraft2012,Avvisati2015}.
Particles with more complex patch architectures and low coordination numbers enable open crystalline
structures, including diamond-like
lattices with relevance for photonic
applications~\cite{Morphew2018,Teixeira2017}.
These systems can be rationalised based on anisotropy and valence
concepts~\cite{Glotzer2007,Hueckel2021}.
A particularly striking consequence of particles having low valence is the
emergence of thermodynamically stable empty liquids
at vanishing packing fraction~\cite{Bianchi2006}, which
evolve upon cooling into equilibrium gel networks stabilised
by long-lived bonds---networks that neither undergo phase
separation nor coarsen or age with
time~\cite{Sciortino2017,Ruzicka2011,Russo2022}.


The richness of three-dimensional (3D) behaviour naturally raises
the question of how geometric confinement modifies
self-assembly as dimensionality is reduced.
Restricting patchy particles to planar geometries enhances
bond directionality and stabilises open networks depending sensitively on patch number, coverage,
and interaction
range~\cite{Doppelbauer2010,Sato2021,Jonas2022,Gharibi2024}.
In two dimensions the interplay between patch geometry and
interaction range selects structures distinct from 3D bulk
phases: triblock Janus particles confined to
quasi-two-dimensional monolayers form kagome lattices, while
three- and four-patch particles self-assemble into honeycomb
and square lattices,
respectively~\cite{Chen2011,Gharibi2024}.
Due to the dimensional restriction, anomalous chain length distribution with a high monomer concentration arises in some divalent patchy particle systems ~\cite{Jonas2022}.
 
An even more extreme form of geometric frustration arises in
quasi-one-dimensional (q1D) channels, where the width is
reduced below two particle diameters, enforcing single-file
motion in which each particle interacts exclusively with its
nearest neighbours while retaining full rotational
freedom~\cite{Lebowitz1983,Marshall2015,Gurin2022}.
Bond formation then requires strict angular alignment,
making the bonding probability a sharply orientation-dependent
quantity that couples rotational and translational degrees of
freedom in a manner with no bulk analogue.
In these settings, Janus particles form helical chains with
tunable pitch, while rod-like particles exhibit anomalous
phase behaviour arising from the competition between
end-to-end bonding and perpendicular
stacking~\cite{Fernandez2015,Gurin2022,Gurin2025}.
From a theoretical point of view, the unique appeal of q1D and
strictly one-dimensional systems stems from the fact that the
transfer operator method provides exact results,
which makes it possible to test the reliability of the approximate theories
widely used for higher-dimensional
systems~\cite{Herzfeld1934,Montero2019,Fantoni2021,Vo2003,Fantoni2017}.


The standard method for associating and patchy systems is
the well-known Wertheim's first-order thermodynamic
perturbation theory (WTPT1), which is formulated as a graph
expansion of the grand partition function with the fractions
of sites not in bond~\cite{Wertheim1984a,Wertheim1984b,
Wertheim1986a,Wertheim1986b}.
In the formalism, the Helmholtz free energy separates into
a reference hard-body contribution and an association
contribution~\cite{Wertheim1984b,Wertheim1987}.
A central quantity is the fraction of particles at
site $i$ not engaged in a bond ($X^i_{u}$), determined self-consistently
through a mass action law involving the pair
distribution function of the reference hard body 
system and the Mayer function of the
associative potential~\cite{Wertheim1984a,Wertheim1984b,
Wertheim1986a,Wertheim1987}.
Within WTPT1 under
single-bonding constraints, the association free energy
reduces to the compact form
$\sum_i \bigl[\ln X^i_{u} - X^i_{u}/2 + 1/2\bigr]$,
which is zero for no bonds ($X^i_{u}=1$) and less than zero for
bond ($0<X^i_{u}<1$)~\cite{Wertheim1984a,Wertheim1984b,
Wertheim1986a,Wertheim1986b,Wertheim1987}.
This framework underlies the statistical associating fluid
theory (SAFT), which became the method of choice for
phase-equilibrium calculations involving complex
fluids~\cite{Jackson1988,GilVillegas1997,Muller2001,
Economou2002,Lira2022}.
Subsequent extensions have been applied to
hydrogen-bonding molecules, chain and polymer fluids, and a
range of complex fluids including aqueous mixtures and
liquid-crystalline
systems~\cite{Muller2001,McGrother1997,Economou2002}.
Despite the wide range of applications, the original WTPT1 does not take into account the effect of the geometric arrangement of sites on the particle surface,
making it unsuitable for orientationally ordered
assemblies~\cite{Wertheim1984a,Wertheim1984b,Wertheim1986a,
Wertheim1987}.
A further issue comes from the non-spherical
particle's shape, which raises additional difficulties in the calculation of the free energy of patchy particles ~\cite{Kalyuzhnyi2026}. It is an extra problem of WTPT1  that the natural variable of the experiments is the pressure and not the density.
These issues have recently been addressed: WTPT1 has been reformulated in
the isothermal--isobaric (NPT) ensemble by exploiting
exact NVT--NPT equivalence relations for the hard-sphere
reference fluid, yielding thermodynamically consistent NPT
expressions for the unbonded fractions, the compressibility
factor, and the internal energy~\cite{Alsaifi2025}; and
TPT has been combined with the interaction-site
model formalism to bypass the angular integrals, enabling
accurate treatment of phase behaviour and percolation in
suspensions of anisotropic platelet colloids decorated with
directional bonding sites~\cite{Kalyuzhnyi2026}.


In the present work, we address how the geometry of the
particle shape and the selectivity of the patch--patch
interaction together determine the orientational order, the
fraction of unbonded sites, and the equation of state of a
q1D associating fluid.
To achieve this, we investigate fluids of hard bodies
carrying two chemically distinct A and B patches interacting
through a Kern--Frenkel
potential~\cite{Gurin2026,Gurin2025}.
We choose a superellipse shape interpolating 
between rectangles and ellipses through a shape exponent and
an aspect ratio, allowing a systematic isolation of the roles
of particle elongation and corner sharpness.
We show that the directional patch--patch association
actively promotes horizontal alignment and network growth
along the q1D channel, whereas hard-body exclusion forces a
realignment toward vertical ordering.
The strong competition between these two factors results in a
pronounced structural crossover between a horizontally
ordered structure---where particles are self-assembled into
long chains---and a vertically ordered structure, where the
particles stand up to optimise packing and become entirely
unbonded.
Because this structural change is so abrupt and intense under
strong confinement, it creates severe kinetic bottlenecks
during compression.
We generalize WTPT1 by making the fraction of sites not in a
bond explicitly orientation-dependent.
By combining the extended formulation with the Parsons--Lee
theory of hard bodies (PL+WTPT1), we provide a
reliable perturbation theory for the phase behaviour of
associating fluids.



\section{\ Model}
\label{sec:model}

   We consider a q1D system of $N$ hard
   particles with superellipse shape, whose centers are confined to
   move along a straight line coinciding with the $x$-axis, while the
   particles retain continuous rotational freedom in the
   $xy$-plane. The configuration of particle $i$ is thus fully
   specified by the position $x_{i}$ of its center and its orientation
   angle $\varphi_{i}$. In a coordinate system fixed to the particles
   $(x^{\prime},y^{\prime})$, the boundary of a superellipse is
   defined by:
\begin{equation}
   \left|\frac{x^{\prime}}{a}\right|^{n}+\left|\frac{y^{\prime}}{b}\right|^{n}=1,
 \label{eq:superellipse}
\end{equation}
   where $a$ and $b$ denote the semi-axes of the particle, and $n$ is
   the shape exponent controlling the sharpness of the particle
   corners. In this study, we assume $b > a$, corresponding to
   a particle elongated along its major axis ($y'$-axis). The aspect
   ratio is defined as $k = b/a$ ,i.e., $k>1$. The special case $n = 2$ reduces to
   an ordinary ellipse, whereas $n \to \infty$ yields a hard
   rectangle. The orientation of a particle is described by the unit
   vector $\mathbf{\omega} = (-\sin \varphi, \cos \varphi)$, which
   points along its principal axis. Here, $\varphi$ is measured with
   respect to the $y$-axis, such that $\varphi = 0$ corresponds to
   vertically aligned particles (major axis perpendicular to the
   channel axis) and $\varphi = \pm\pi/2$ corresponds to perfect
   horizontal order along the channel. Transforming
   Eq.~(\ref{eq:superellipse}) into the original reference frame, the
   boundary of a superellipse with orientation $\mathbf{\omega}$ satisfies
\begin{equation}
  \label{superellipse}
   \left|\frac{x \cos\varphi + y
       \sin\varphi}{a}\right|^{n}+\left|\frac{-x \sin\varphi + y
     \cos\varphi}{b}\right|^{n}=1.
\end{equation}

   The  pair interaction between particles $i$ and $j$ is the sum of hard-core repulsion ($U_{\text{HSE}}$) and short-ranged directional attraction ($U_{\text{bond}}$)
\begin{equation}
  \label{U}
U(x_{ij},\varphi_{i},\varphi_{j}) = U_{\text{HSE}}(x_{ij},\varphi_i,\varphi_j) + U_{\text{bond}}(x_{ij},\varphi_i,\varphi_j) ,
\end{equation}
   where $x_{ij} = |x_{i} - x_{j}|$ is the center-to-center 
   distance. In the case of superellipses $U_{\text{HSE}}$ is given by 
   
 \begin{equation}
  \label{U_HSE}
  U_{\text{HSE}}(x,\varphi,\varphi')
  = \begin{cases}
    \infty & \text{if } x<\sigma(\varphi,\varphi')
    \\
    0    & \text{otherwise,}
  \end{cases}
\end{equation}
   where $x$ is the distance and $\sigma(\varphi,\varphi')$ is the
   contact distance between two neighbouring particles with
   orientations $\varphi$ and $\varphi'$. Our procedure for determining the contact distance between two superellipses is presented in~\cite{mizani2025emergence}.

\begin{figure}[!ht]
  \centering
  \includegraphics[scale=.28]{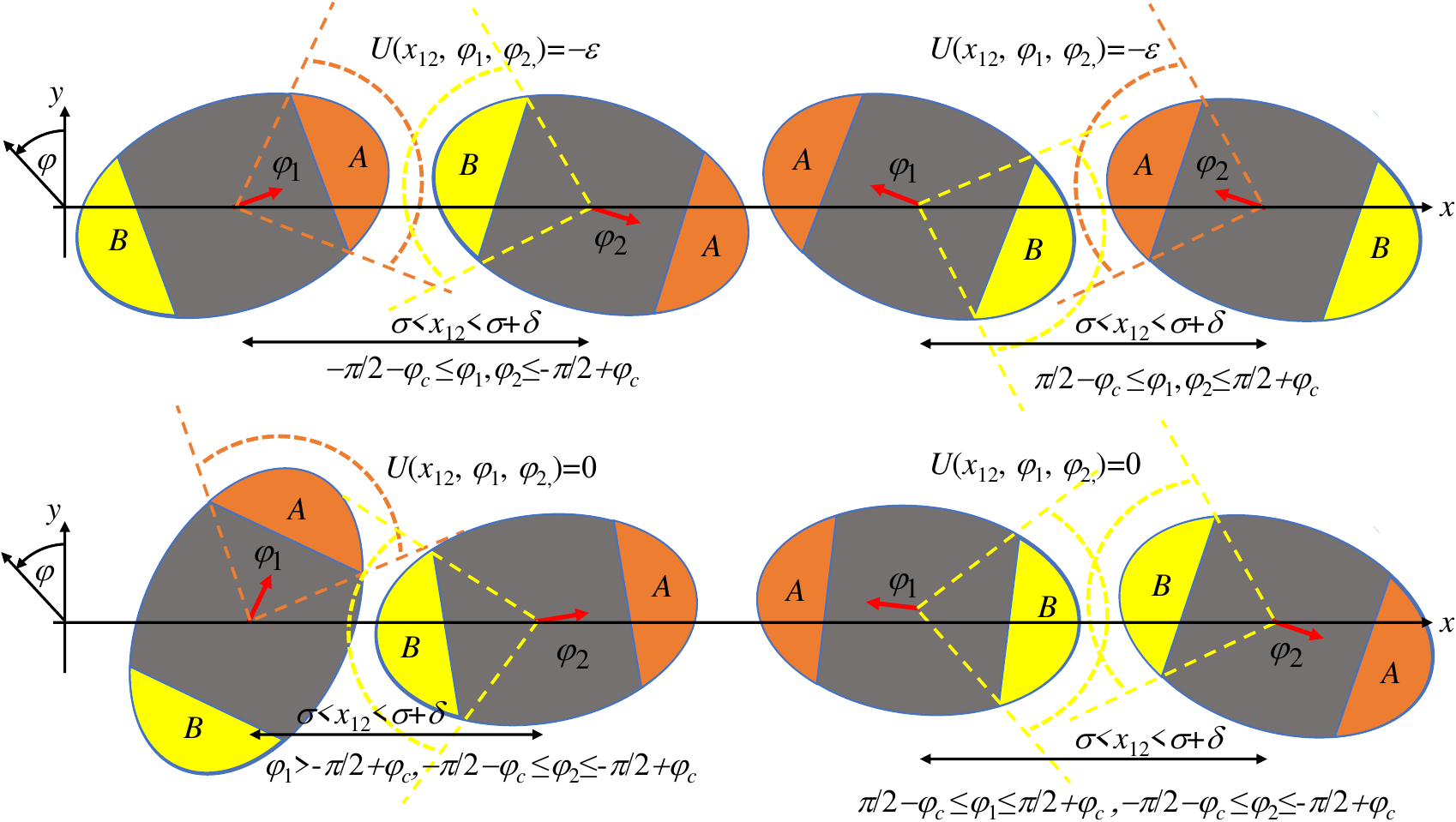}
  \caption{Kern--Frenkel patch--patch interaction between A and B
    sites.  The figure illustrates bonded and non-bonded
    configurations.}
 \label{fig:model}
\end{figure}

    Regarding short-range attraction, each particle
    is decorated with two chemically distinct bonding sites, labeled
    by $A$ and $B$, positioned at opposite tips along the particle's
    major axis. The selective attraction is governed by a 
    Kern-Frenkel pair potential, where a bond forms only between the $A$
    and $B$ patches of neighboring particles ($A\text{–}B$ bond),
    while $A\text{–}A$ and $B\text{–}B$ bonds are
    forbidden. Furthermore, each patch can form at most one bond. A
    bond  energy ($-\epsilon$) is established if the neighbouring  particles satisfy
    both a distance constraint and an angular condition shown in Fig.~\ref{fig:model}. Therefore,
\begin{align}
  \label{Kern-Frenkel} 
    U_{\text{bond}}^{AA}(x,\varphi,\varphi')
 &= U_{\text{bond}}^{BB}(x,\varphi,\varphi')
  = 0 ,
\nonumber\\
    U_{\text{bond}}^{AB}(x,\varphi,\varphi')
 &= 
   \begin{cases}
    -\epsilon, & \text{if } \varphi,\varphi' \in I_- \text{ and}\\
               & \quad\sigma(\varphi,\varphi')<x<\sigma(\varphi,\varphi')+\delta
    \\[3pt]
    0          & \text{otherwise,}
  \end{cases}
\nonumber\\
    U_{\text{bond}}^{BA}(x,\varphi,\varphi')
 &= 
   \begin{cases}
    -\epsilon, & \text{if } \varphi,\varphi' \in I_+ \text{ and}\\
               & \quad\sigma(\varphi,\varphi')<x<\sigma(\varphi,\varphi')+\delta
    \\[3pt]
    0          & \text{otherwise,}
  \end{cases}
\end{align}
   where $\delta$ is the attraction range, $\varphi_{c}$ is the half of the bonding angle, and
$I_{\pm}$ denotes the angle interval of the
   two possible A--B bonds as follows:
\begin{equation}
  \label{I}
  I_-=[-\frac{\pi}{2}-\varphi_{c},-\frac{\pi}{2}+\varphi_{c}],
  \quad
  I_+=[\frac{\pi}{2}-\varphi_{c},\frac{\pi}{2}+\varphi_{c}].
\end{equation}
 In our 
 present work, we study systems where  $\varphi_c=20^\circ$ and $\delta=0.2$ measured
 in the unit of shortest length ($2a$). The total bonding energy is the sum of the energies of the possible bonds, i.e.,
\begin{equation}
  \label{U_bond}
  U_{\text{bond}}(x,\varphi,\varphi')
  = \sum_{k,l\in\{A,B\}} U_{\text{bond}}^{kl}(x,\varphi,\varphi') .
\end{equation}
    As an illustration in Fig.~\ref{fig:model} we present two cases where the bonding conditions are satisfied and other two cases where not.
   We can see that the two geometric criteria are satisfied in the upper panel while the angle criterion is violated in the lower panel.

\begin{figure}[!ht]
  \centering
  \includegraphics[scale=.33]{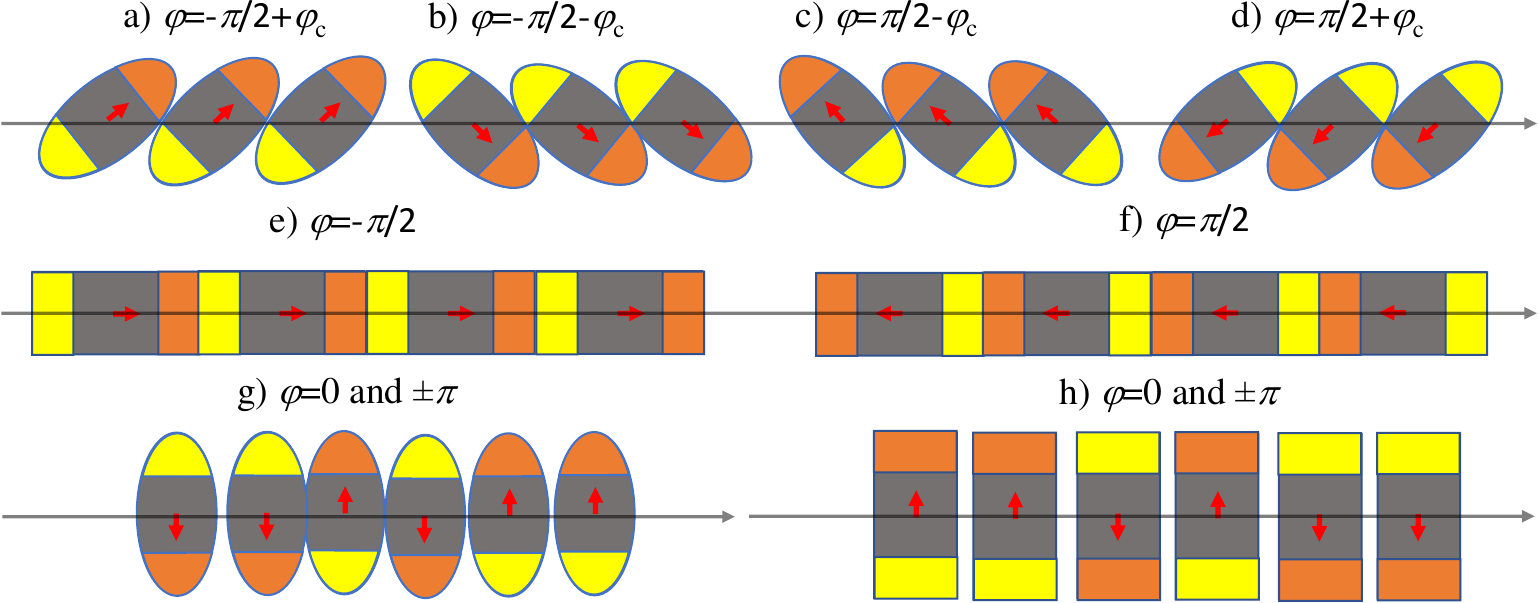}
  \caption{Closely packed structures of patchy hard ellipses and patchy hard
    rectangles. Panels (a–d) correspond to tilted close packing structures of ellipses, panels (e–f) to horizontal close packing structures of rectangles and panels (g–h) show the vertical close-packing structures of ellipses and rectangles.}
 \label{fig:model2}
\end{figure}

   The intricate coupling between particle geometry, corner sharpness,
   and patch alignment leads to distinct high pressure structures,
   which are summarized in Fig.~\ref{fig:model2}. As the pressure
   goes to infinity, the close packing configuration for any aspect
   ratio $k > 1$ corresponds to a perfectly vertical arrangement
   ($\varphi = 0$ or $\varphi=\pm\pi$). This structure minimizes the excluded distance
   along the confinement axis, thereby maximizing the packing
   efficiency of the hard cores. When strong associative $A\text{–}B$
   attractions are present, however, a sharp structural competition
   emerges between bonding optimization and packing optimization at
   high pressure. This competition depends fundamentally on the shape
   exponent $n$.

   In the case of patchy hard ellipses, $n = 2$, as illustrated in
   Figs.~\ref{fig:model2}a–d, the smoothly curved profile of the ellipses
   allows the particles to form a highly dense, tilted chain
   structure at high pressures. By adopting a
   cooperative finite tilt angle, the ellipses maximize the number of
   active $A\text{–}B$ bonds while simultaneously minimizing the
   occupied length along the channel axis. This bonded and tilted
   configuration can remain stable at most up to an intermediate boundary density
   $\rho_b$, where two neighboring particles are in direct contact at
   the maximum threshold of the patch opening angle, i.e.,
\begin{equation}
    \rho_b = \frac{1}{2} \sqrt{ \frac{\cos^2\varphi_c}{b^2} +
             \frac{\sin^2\varphi_c}{a^2} }.
\end{equation}
   This threshold number density can be achieved by any of the four
   equivalent particle orientations illustrated in
   Fig.~\ref{fig:model2}a–d. Above $\rho_b$, the spatial
   limitation makes the bonded, tilted structure impossible. Consequently, the $A\text{–}B$ bonds, break
   and the particles undergo a structural crossover into the above
   mentioned vertically ordered phase preferred by the hard-body
   interactions (Fig.~\ref{fig:model2}g). Patchy ellipses thus exhibit
   a distinct tilted--vertical structural competition.

   In contrast, in the case of patchy hard rectangles, $n \to \infty$,
   the presence of perfectly straight edges and right-angled corners
   completely alters the bonding landscape. As shown in
   Figs.~\ref{fig:model2} e-f, the high pressure structure that
   preserves all $A\text{–}B$ bonds is strictly horizontal
   ($\varphi = \pm\pi/2$), where the flat ends fit together, and
   particles form a straight chain. This perfectly horizontal bonded
   phase can remain stable at most up to an intermediate boundary number density given by
\begin{equation}
   \rho_b = \frac{1}{2b},
 \label{eq:rhoc}   
\end{equation}
    which corresponds to the reciprocal of the contact distance along
    the channel for a purely horizontal rectangle. Beyond this
    boundary density, it becomes geometrically impossible to pack the
    particles horizontally. To accommodate further compression, the
    rectangles must stand up to optimize packing fraction, completely
    breaking the $A\text{–}B$ bonds and entering a vertically ordered
    state ($\varphi = 0$ or $\varphi=\pm\pi$, Fig.~\ref{fig:model2}h). Patchy rectangles therefore
    undergo a horizontal–vertical structural competition.

In fact, the above described ellipse-like behavior also persists for $n>2$, and the rectangle-like behavior also persists for $n<\infty$. The value of the deformation parameter $n$ that separates the two behaviors depends on the value of the bonding half-angle $\varphi_c$.
\section{\ Theory}
\subsection{Transfer Operator Method}
\label{sec:TOM}

   The transfer operator method (TOM), which can be applied when the
   system is q1D and the pair interaction is
   short-ranged, provides exact thermodynamic and structural
   properties for q1D fluids. Here we give a short summary of this
   method.

   The configuration part of the isobaric partition function, which depends
   on the number of particles ($N$), the pressure ($P$) and the temperature ($T$), can be written as
\begin{align}
     Z_N(P, T)
  =&
   \frac{1}{N!}
      \int\limits_0^\infty dL\; e^{- \beta P L} \times
\nonumber\\
    & \int\limits_{-\pi}^{\pi}  \Bigl( \prod_{i=1}^N d  \varphi_i \Bigr)
     \int\limits_0^{L}     \Bigl( \prod_{i=1}^N d  x_i \Bigr) 
     e^{-\beta \sum_{i<j} U(x_{ij},\varphi_i,\varphi_j) }
 ,
 \label{eq:Z_def}
\end{align}
   where $\beta^{-1}=k_B T$, $L$ is the length of the system along the
   $x$ axis, and the particle $i$ has $x_i$ position and $\varphi_i$ orientation. Restricting the domain of the integration in Eq.(\ref{eq:Z_def})
   for a fixed order of the particles, $x_1 \leq \dots \leq x_N$, we
   can omit the prefactor $1/N!$ because the interparticle potential
   is symmetric under the permutation of particles. Furthermore we
   impose periodic boundary conditions such that $x_{N+1}=x_1$,
   therefore we have $L = \sum_{i=1}^N x_{i,i+1}$, where $ x_{i,i+1}= x_{i+1}- x_{i}$. Due to the fact a given particle interacts only with its left and right neighours, using $ x_{i,i+1}$ instead of $x_i$,
   the $N$-fold integral becomes a product in Eq.(\ref{eq:Z_def}), and $Z_N$ can be expressed as
   \begin{equation}
\label{eq:Tr}
     Z_N(P, T)
  =  \int\limits_{-\pi}^{\pi}  \Bigl( \prod_{i=1}^N d  \varphi_i \Bigr)
     \Bigl( \prod_{i=1}^N K(\varphi_i,\varphi_{i+1}) \Bigr)
   , 
\end{equation}
where
\begin{align}
\label{eq:kernel}
     K(\varphi,\varphi')
  &= \int_{0}^{\infty}
    e^{-\beta P x}\,
    e^{-\beta U(x,\varphi,\varphi')}
    \,dx\nonumber\\
&  = \int_{\sigma(\varphi,\varphi')}^{\infty}
    e^{-\beta P x}\,
    e^{-\beta U_{\text{bond}}(x,\varphi,\varphi')}
    \,dx.
\end{align}
In Eq.~(\ref{eq:kernel}) the lower integration limit,
   $\sigma(\varphi,\varphi')$, ensures that the hard-core repulsion is taken into account. As the
   trace operation is independent of the basis used,
   according to the
   Perron--Frobenius--Jentzsch theorem \cite{Cuesta-Sanchez_JStatPhys_2004}, in thermodynamical limit ($N, L \rightarrow
   \infty$ but $\rho=N/L$ is fixed) the Gibbs free energy   can be written as
\begin{eqnarray}
\label{eq:G}
     G
  =  -N k T \ln \lambda_0,
\end{eqnarray}
where $\lambda_0$ is the largest eigenvalue of the following  eigenvalue problem
\begin{equation}
  \int_{-\pi}^{\pi}
  K(\varphi,\varphi')\,\psi_k(\varphi')\,d\varphi'
  = \lambda_k \psi_k(\varphi).
  \label{eq:eigen}
\end{equation}
   In this equation, 
   $\lambda_k$ is an eigenvalue and $\psi_k(\varphi)$ is the corresponding eigenfunction with the condition of $\lambda_0> |\lambda_1|>|\lambda_2|>...$.
   The equation of state, which relates density to pressure,
   follows from the standard thermodynamic relation
\begin{equation}
  \frac{1}{\rho}
  = \frac{\partial(\beta G/N)}{\partial(\beta P)}
  = -\frac{\partial \log\lambda_0}{\partial(\beta P)}.
  \label{eq:EOS_TOM}
\end{equation}
   For models defined in Sec.~\ref{sec:model}, the integral in
   Eq.~\eqref{eq:kernel} can be evaluated analytically and we obtain
   the following result
\begin{equation}
  K(\varphi,\varphi')
  =
  \begin{cases}
    \dfrac{e^{-\beta P\sigma(\varphi,\varphi')}}{\beta P}
    \Bigl[1+f_M(1 - e^{-\beta P\delta})

    \Bigr],
    \\
 \qquad\quad\text{if }\; \varphi,\varphi'\in I_- \;\text{ or }\;
                   \varphi,\varphi'\in I_+ 
    \\[5pt]
    \dfrac{e^{-\beta P\sigma(\varphi,\varphi')}}{\beta P},
\qquad\qquad
\text{otherwise.}
  \end{cases}
  \label{eq:kernel_12}
\end{equation}
   where $f_M=e^{\beta \varepsilon}-1$ is the Mayer function of the bond. Note that the $f_M(1 - e^{-\beta P\delta})$ term is due to the A--B bond as it becomes 0 for $\varepsilon=0$ and the kernel reduces to the pure
   hard-body expression.

   The orientational
   distribution function (ODF) of particles can be obtained using $\psi_0$ as
\begin{equation}
  f(\varphi) = \psi_0(\varphi)^2 .
  \label{eq:ODF_TOM}
\end{equation}
   By definition, the ODF is normalised as
\begin{equation}
  \int_{-\pi}^{\pi} f(\varphi)\,d\varphi = 1 .
  \label{eq:norm}
\end{equation}
   Note that Eq.~\eqref{eq:ODF_TOM} is due to the symmetry property of Kernel, given by $K(\varphi,\varphi') =
   K(\varphi',\varphi)$.
   
   The nematic order parameter, which quantifies the degree of
   orientational alignment, is defined as
\begin{equation}
  S = \langle \cos 2\varphi \rangle
    = \int_{-\pi}^{\pi} f(\varphi)\cos(2\varphi)\,d\varphi.
  \label{eq:S}
\end{equation}
   It can be shown that $S = 0$ in the isotropic phase, $S > 0$ in vertical
   ordering (particles aligned perpendicular to the channel), and $S <
   0$ in horizontal ordering (particles aligned along the
   channel).

   To obtain the fraction of sites not in bond from the TOM, we need the probability density of finding two
   neighbouring particles at distance $x$ with orientations
   $\varphi$ and $\varphi'$. It is shown in~\cite{Gurin2025} that
\begin{equation}
  f(x,\varphi,\varphi')
  = \frac{\psi_0(\varphi)
      e^{-\beta U(x,\varphi,\varphi')-\beta P x}
      \psi_0(\varphi')
    }{\lambda_0}.
  \label{eq:joint}
\end{equation}
   By integrating Eq.~\eqref{eq:joint} over all distances beyond the
   maximum bonding range of sites $k$ and $l$, one
   obtains the fraction of site $k$ not bonded to the right neighbour as follows
\begin{equation}
  X^{k}_{u,R}(\varphi)
  = \frac{1}{f(\varphi)}
      \sum_{l}
      \int_{-\pi}^{\pi}
      \int_{x^{kl}_{\max}(\varphi,\varphi')}^{\infty}
      f(x,\varphi,\varphi')\,dx\,d\varphi'
    .
  \label{eq:XuR}
\end{equation}
   Here the maximum bonding range, $x^{kl}_{\max}(\varphi,\varphi')$,
  denotes the center-to-center distance between two particles when the
  $k$ site of the particle on the left with orientation   $\varphi$ and the $l$ site of the
  particle on the right with orientation   $\varphi'$ are exactly outside of the interaction range
  (but they would interact with each other if they were closer). The quantity $x^{kl}_{\max}(\varphi,\varphi')$ takes into account
  only the interaction between the $k$ and $l$ sites, regardless of
  the interactions between the other sites.  The fraction of site $k$
  not bonded to the left neighbour, $X^{k}_{u,L}(\varphi)$, is
  obtained analogously by exchanging the roles of $\varphi$ and
  $\varphi'$ in Eq.~\eqref{eq:joint}. 
  If site $k$ bonds to a right (left) neighbor, it can not bond to the left (right) one, therefore $X^{k}_{u,R}(\varphi)$<1 ($X^{k}_{u,R}(\varphi)$=1) and $X^{k}_{u,L}(\varphi)$=1 ($X^{k}_{u,L}(\varphi)$<1).
  These properties of the fraction of site $k$ not in bond can be unified into the following equation
  
\begin{equation}
  X^{k}_{u}(\varphi)
  = X^{k}_{u,L}(\varphi)\,X^{k}_{u,R}(\varphi).
  \label{eq:Xu_TOM}
\end{equation}

   When we discretize the possible values
   of $\varphi$ in Eq.~(\ref{eq:kernel_12}) we get a transfer
   matrix. Although this is an approximation, as the discretization is refined, the results converge to numerically exact values for the thermodynamic and structural properties of the system. Thus, the TOM results serve as exact benchmarks for evaluating the precision of the Parsons–Lee and Wertheim theories.

\subsection{Parsons-Lee Approximation for Hard Superellipses}
\label{sec:PL}

   As the reference system for patchy particles is the system of hard superellipses, we resort to the well-known Parsons–Lee (PL) theory of hard body fluids~\cite{Lee1987,Parsons1979}. 
   In this formalism, the Helmholtz free energy density
   is given by
\begin{align}
  \frac{\beta F_{\text{HSE}}}{N}
  =& \ln\rho - 1+ \int_{-\pi}^{\pi} f(\varphi)\ln\!\bigl(f(\varphi)\bigr)\,d\varphi
\nonumber\\
 &   - \frac{\ln\bigl(1 - \rho\davg\bigr)}{\davg}\,\savg.
  \label{eq:F_PL}
\end{align}
   The first two
   terms of Eq.~\eqref{eq:F_PL} represent the translational (or
   ideal-gas) contribution, the third is the orientational
   entropy term, and the last one accounts for the hard-body
   excluded-volume interactions. Note that only the ideal gas and the orientational entropy terms are exact, while the hard body contribution is approximate as the hard superellipses are mapped into the hard disks. Instead of equating the area of the superellipse with that of a disk ~\cite{Lee1988}, we identify the diameter of the disk as the orientational average of the occupied length of the superellipse along the $x$ axis. Therefore we use
\begin{equation}
  \davg = \int_{-\pi}^{\pi} f(\varphi)\,d(\varphi)\,d\varphi,
  \label{eq:davg}
\end{equation}
and
\begin{equation}
  \savg
  = \int_{-\pi}^{\pi}\!\int_{-\pi}^{\pi}
    f(\varphi)\,f(\varphi')\,\sigma(\varphi,\varphi')
    \,d\varphi\,d\varphi',
  \label{eq:savg}
\end{equation}
   where $d(\varphi)$ is the occupied length of a superellipse particle with
   orientation $\varphi$ along the $x$ axis.

   The equilibrium ODF is obtained by minimising Eq.~\eqref{eq:F_PL}
   with the normalisation constraint of
   Eq.~\eqref{eq:norm}, which results in the following self-consistent equation
\begin{align}
  f(\varphi) = \frac{1}{C}
  \exp\!&\left[
    \frac{2\ln(1-\rho\davg)}{\davg}
    \int_{-\pi}^{\pi}
    f(\varphi')\,\sigma(\varphi,\varphi')\,d\varphi'
    \right.\nonumber\\&\;\;\left. 
    - \frac{d(\varphi)\,\savg}{\davg}
      \Bigl(\frac{\rho}{1-\rho\davg} + \frac{\ln(1-\rho\davg)}{\davg}
      \Bigr)
  \right],
  \label{eq:ODF_hard}
\end{align}
   where $C$ is the angular integral
   of the exponential function of Eq~\eqref{eq:ODF_hard}. Using $\beta P =
   \rho^{2}\,\partial(\beta F/N)/\partial\rho$ and Eq.~\eqref{eq:F_PL},
   the pressure of the hard superellipse system is given by
\begin{equation}
  \beta P_{\mathrm{HSE}} = \rho + \frac{\rho^{2}\savg}{1-\rho\davg}.
  \label{eq:P_hard}
\end{equation}

\subsection{Wertheim First-Order thermodynamic perturbation theory}
\label{sec:Wertheim}

   To account for the A--B bonding between patchy superellipses, we
   generalise the original Wertheim first-order thermodynamic perturbation
   theory (WTPT1) and combine it with the PL theory of the reference hard body 
   system. Considering $X^{k}_{u}(\varphi)$ as a new key quantity of the original Wertheim's TPT1 approach, we assume that the bond contribution of the Helmholtz free energy is a functional of both $X^{k}_{u}(\varphi)$ and $f(\varphi)$ as follows
\begin{align}
  \frac{\beta F_{\text{Bond}}}{N}
  =&  \sum_{k\in\{A,B\}}\int_{-\pi}^{\pi} f(\varphi)
     \Bigl[\ln X^{k}_{u}(\varphi)
           - X^{k}_{u}(\varphi) + 1\Bigr]\,d\varphi
   \notag\\
  &
   - \frac{\rho}{2}
     \sum_{k,l\in\{A,B\}}
     \int_{-\pi}^{\pi}\!\int_{-\pi}^{\pi}
     f(\varphi)\,f(\varphi') \times
 \nonumber\\
 &\qquad\qquad\qquad X^{k}_{u}(\varphi)X^{l}_{u}(\varphi')
     \Delta^{kl}(\varphi,\varphi')
     d\varphi d\varphi'.
  \label{eq:F_full}
\end{align}
In this equation, the bonding integral is given by
\begin{equation}
\label{Delta}
     \Delta^{kl}(\varphi,\varphi')
  =\! \int_{-\infty}^\infty\!\!\!\! g_{\mathrm{HSE}}(x,\varphi,\varphi')(e^{-\beta U_{\text{bond}}^{kl}(x,\varphi,\varphi')}-1)\, dx,
\end{equation}
   where 
   $g_{\mathrm{HSE}}$  is the pair correlation function of
   the reference 1D hard superellipse fluid. Note that the sum of  Eq.~\eqref{eq:F_PL} and  Eq.~\eqref{eq:F_full}  constitutes the total free energy of patchy hard superellipses, i.e., 
   ${F}= F_{\text{HSE}}+ F_{\text{Bond}}$. As $g_{\mathrm{HSE}}$ is not analytic, we approximate it with the result of the mapping procedure between hard spheres and hard superellipses suggested by Parsons ~\cite{Parsons1979}, which assumes that $g_{\mathrm{HSE}}$$\approx $$g_{\mathrm{HS}}$, where $g_{\mathrm{HS}}$ is the pair correlation function of 1D hard spheres having $\davg$ diameters. As the range of the site--site interaction, $\delta$, is very short, $g_{\mathrm{HS}}$ can be
   approximated with its contact value,
   i.e., $g_{\mathrm{HSE}}\approx\frac{1}{1 - \rho\davg}$. Thus, for
   the Kern--Frenkel site--site interaction (see
   Eq.~(\ref{Kern-Frenkel})) we get the following analytic expressions for the bonding integrals
\begin{align}
  \Delta^{AB}&(\varphi,\varphi')
  =
  \Delta^{BA}(\varphi,\varphi')=\nonumber\\
&  =
  \begin{cases}
    \displaystyle
    f_{M}\,\frac{\delta}{1 - \rho\davg}, &
      \text{if }\; \varphi,\varphi'\in I_- \;\text{ or }\;
                   \varphi,\varphi'\in I_+  \\
    0, & \text{otherwise,}
  \end{cases}
  \label{eq:Delta}
\end{align} 
   and
\begin{equation}
  \Delta^{AA}(\varphi,\varphi')=\Delta^{BB}(\varphi,\varphi')=0 .
  \label{Delta_AA-BB}
\end{equation}

   The equilibrium $X^{k}_{u}(\varphi)$,
   can be obtained by functional minimization of the bond free energy contribution (Eq.~\eqref{eq:F_full})
   with respect to $X^{k}_{u}(\varphi)$. From $\delta(\beta
   F/N)/\delta X^{k}_{u}(\varphi) = 0$ and using the symmetry of
   $\Delta^{kl}(\varphi,\varphi') =
   \Delta^{lk}(\varphi',\varphi)$, we end up with the orientation-dependent law of mass action given by
\begin{equation}
  \frac{1}{X^{k}_{u}(\varphi)}
  = 1 + \rho\sum_{l\in\{A,B\}}
        \int_{-\pi}^{\pi}
        f(\varphi')\,X^{l}_{u}(\varphi')\,
        \Delta^{kl}(\varphi,\varphi')\,d\varphi'.
  \label{eq:LMA}
\end{equation}
With the help of this law, it can be shown that Eq.~\eqref{eq:F_full} simplifies into the well-known form of the original Wertheim's bond term by performing additional orientational average as follows

\begin{align}
  \frac{\beta F_{\text{Bond}}}{N}
  =&  \sum_{k\in\{A,B\}}\int_{-\pi}^{\pi} f(\varphi)
     \Bigl[\ln X^{k}_{u}(\varphi)
           - \frac{X^{k}_{u}(\varphi)}{2} + \frac{1}{2}\Bigr]\,d\varphi.
   \notag\\
  &
.
  \label{eq:F_bond}
\end{align}
 As should be, this bond contribution is non-zero for bonds ($X^{k}_{u}(\varphi)$<1), but is exactly zero in the lack of bonds ($X^{k}_{u}(\varphi)$=1).
   Using Eqs.~(\ref{eq:Delta}-\ref{Delta_AA-BB}), the orientation dependence of $X_{u}^k(\varphi)$ simplifies to
\begin{equation}
  X^{A}_{u}(\varphi)=X^{B}_{u}(\varphi) 
  =
  \begin{cases}
    X_{u}, & \varphi\in I_+ \text{ or } \varphi\in I_- \\
    1,              & \text{otherwise.}
  \end{cases}
  \label{eq:Xu_12}
\end{equation}
 Substituting this result into Eqs.~(\ref{eq:LMA}) the law of mass action becomes very simple as
\begin{equation}
  \frac{1}{X_{u}}
  = 1
  + \frac{f_{M}\,\delta\,\rho X_{u}}{1-\rho\davg}
    \int_{\pi/2-\varphi_{c}}^{\pi/2+\varphi_{c}}
    f(\varphi)\,d\varphi .
  \label{eq:LMA_12}
\end{equation}
   Note that this simple equation relies on the assumptions made for the bonding integral and
   $f(\varphi)=f(\varphi+\pi)$ is also used.
   The equilibrium ODF is obtained by
   minimising the total free energy with
   respect to $f$ with the normalization constraint
   of Eq.~\eqref{eq:norm}.  This minimization procedure requires the functional
   derivatives of $\Delta$ with respect to $f$ and with respect
   to $\rho$, which are given by
\begin{equation}
  \frac{\delta \Delta^{kl}(\varphi,\varphi')}{\delta f(\varphi)}
  = \frac{d(\varphi)\rho}{1-\rho\davg}
    \,\Delta^{kl}(\varphi,\varphi'),
  \label{eq:dDdf}
\end{equation}
\begin{equation}
  \frac{d\,\Delta^{kl}(\varphi,\varphi')}{d\rho}
  = \frac{\davg}{1-\rho\davg}
    \,\Delta^{kl}(\varphi,\varphi').
  \label{eq:dDdrho}
\end{equation}
   Using these equations the minimization procedure results in the following self-consistent equation for the equilibrium ODF
\begin{align}
  f(\varphi)
 & = \frac{1}{c}
  \frac{1}{\displaystyle{\prod_{k\in\{A,B\}}}X^{k}_{u}(\varphi)}
  \exp\Biggl[\nonumber\\
   & \frac{2\ln(1-\rho\davg)}{\davg}
    \int_{-\pi}^{\pi}
    f(\varphi')\,\sigma(\varphi,\varphi')\,d\varphi'
  \nonumber\\
  &
  - \frac{d(\varphi)\,\savg}{\davg}
      \left(\frac{\rho}{1-\rho\davg} + \frac{\ln(1-\rho\davg)}{\davg}\right)
\nonumber\\
      &\;+\;
    \frac{\rho\,d(\varphi)}{1-\rho\davg}
    \frac{1}{2}\sum_{k\in\{A,B\}}
    \int_{-\pi}^{\pi}
    f(\varphi')\bigl(1 - X^{k}_{u}(\varphi')\bigr)\,d\varphi'
    \Biggr].
  \label{eq:ODF_full}
\end{align}
   Here, $c$ is a normalization constant ensuring that Eq.~(\ref{eq:norm}) is satisfied. Using
Eqs.~(\ref{eq:Xu_12}-\ref{eq:LMA_12}), we end up with
\begin{align}
  f(\varphi)
  &= \frac{1}{c}
  \frac{1}{(X_{u})^2}
  \exp\Biggl[\nonumber\\
 &   \frac{2\ln(1-\rho\davg)}{\davg}
    \int_{-\pi}^{\pi}
    f(\varphi')\,\sigma(\varphi,\varphi')\,d\varphi'
  \nonumber\\
  &
  - \frac{d(\varphi)\,\savg}{\davg}
      \left(\frac{\rho}{1-\rho\davg} + \frac{\ln(1-\rho\davg)}{\davg}\right)
\nonumber\\
      &\;+\;
    \frac{\rho\,d(\varphi)}{1-\rho\davg}
    2(1 - X_{u})
    \int_{\pi/2-\varphi_{c}}^{\pi/2+\varphi_{c}}
    f(\varphi')\,d\varphi'
    \Biggr] .
  \label{eq:ODF_12}
\end{align}
 Note that Eq.~\eqref{eq:ODF_12} reduces to Eq.~\eqref{eq:ODF_hard} in the lack of bond ($X_{u}=1$). Having obtained the equilibrium ODF and $X_{u}$, the pressure can be determined using $\beta P =
   \rho^{2}\,\partial(\beta F/N)/\partial\rho$. With the help of Eqs.~\eqref{eq:F_PL},~\eqref{eq:F_full},~\eqref{eq:LMA_12} and~\eqref{eq:dDdrho} the equation of state simplifies to
\begin{equation}
  \beta P
  = \rho + \frac{\rho^{2}\savg}{1-\rho\davg}
    - \frac{\rho(1 - X_{u})}{1-\rho\davg}
      \int_{\pi/2-\varphi_{c}}^{\pi/2+\varphi_{c}}
      f(\varphi)\,d\varphi.
  \label{eq:EOS}
\end{equation}
   The first two terms of Eq.~\eqref{eq:EOS} gives the reference
   hard-body pressure (see Eq.~\eqref{eq:P_hard}), while the third term is a negative bond contribution as $X_{u}\leq 1$ reflecting the reduction in pressure due to chain formation.

   We solve Eqs.~\eqref{eq:LMA_12} and \eqref{eq:ODF_12} simultaneously together with the
    normalization condition (Eq.~\eqref{eq:norm}) for $X_{u}$ and
   $f(\varphi)$ at a given density $\rho$. To do this, we discretise the angle, use trapezoidal quadrature for the integrals and solve the equations with iteration method. Once $X_{u}$ and
   $f(\varphi)$ are known, the pressure (Eq.~\eqref{eq:EOS}) and the
   order parameter (Eq.~\eqref{eq:S}) follow directly.

\section{\ Results}
\label{sec:R}
\subsection{TOM Results for Ellipses and Superellipses}
We first present the results of the transfer operator method for the thermodynamic and structural behavior of q1D patchy superellipses.
With this method, we systematically isolate the effects of
geometric anisotropy---via the aspect ratio ($k$) and shape
exponent ($n$)---and patch-driven association across the
entire density regime.
In the following, we analyse our results alongside two
reference systems to decouple individual contributions:
(1) the $k=1$ systems (patchy hard disks for $n=2$ and
square-like patchy hard superdisks for $n=10$) decorated
with two identical patches with A--B bonding, and
(2) the corresponding unpatched hard-body reference systems
($\varepsilon=0$).
After this we examine the accuracy of the PL 
theory for q1D hard ellipses and that of the PL theory
combined with the extended WTPT1 for q1D patchy
hard ellipses.
We show that PL+WTPT1 provides reasonable results, which
makes it suitable for studying higher-dimensional patchy
hard-body fluids.

\subsubsection{Structural Distribution}
\begin{figure}[!ht]
\centering
\subfigure[]{
\includegraphics[scale=.95]{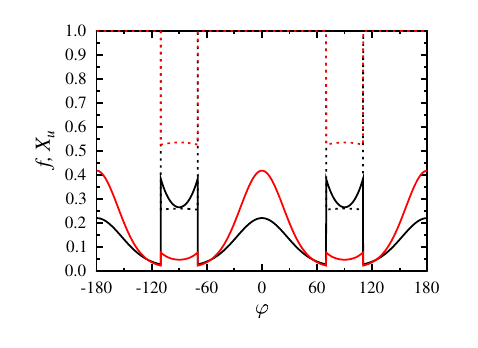}
\label{3a}
}

\subfigure[]{
\includegraphics[scale=.95]{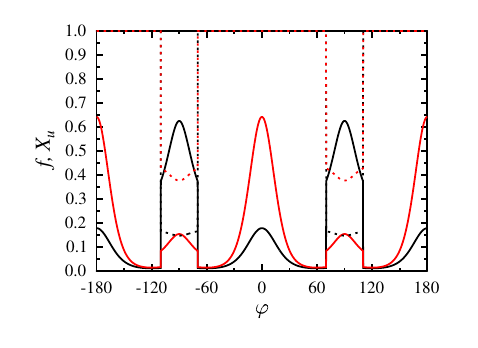}
\label{3b}
}

\caption{Orientational distribution functions, $f(\varphi)$ (solid lines), and fractions of unbonded sites, $X_u(\varphi)$ (dashed lines), as functions of the particle orientation $\varphi$ for $\beta\epsilon=4$, $\delta=0.2$, $\varphi_c=20^\circ$, and $k=1.5$. The upper panel is for ellipses ($n=2$), while the lower one is for rectangles ($n=10$). In each panel, results are shown for reduced pressures $P^{*}=5$ (black) and $P^{*}=7$ (red).
}
\label{fig:odf_xu}
\end{figure}

We start with the structural distribution functions at low
and high pressures to validate the structures predicted in
the Model section.
Fig~\ref{fig:odf_xu} 
displays the orientational
distribution function, $f(\varphi)$, and the
corresponding fraction of sites not in a bond, $X_u$,
providing information about the orientational order and the extent of bonding. We can see that the curves support the predictions for intermediate bonded configurations and the high-pressure
close-packing structure without bonds as illustrated in
Fig~\ref{fig:model2}. In the case of ellipses ($n=2$) at intermediate densities,
$f(\varphi)$ is not peaked at the perfectly horizontal
orientation ($\varphi=\pm\pi/2$, where the major axis lies
along the channel axis).
Instead, the distribution splits and displays prominent
peaks located precisely at $\varphi=\pm\pi/2\pm\varphi_c$.
The same physical signature can be seen in $X_u$, which
reveals that the majority of active bond formations occur at
$\pm\pi/2\pm\varphi_c$.
This provides direct evidence for the tilted close-packed
chains: the ellipses are rotated away from the horizontal
channel axis by $\varphi_c$ to preserve full A--B bonding
connectivity while optimising their packed structure along
the straight line. For rectangle-like superellipses ($n=10$), the scenario is
fundamentally different due to the presence of straight
edges.
As shown in Fig~\ref{fig:odf_xu}, $f(\varphi)$ is
uniquely peaked at the perfectly horizontal orientation
($\varphi=\pm\pi/2$), and the majority of bonds occur
precisely at $\varphi=\pm\pi/2$, minimising $X_u$.
This confirms that the A--B bonds and straight edges of the
rectangles together align the particles to form a linear
chain, locking them into a horizontal configuration and
removing the angular tilt degeneracy seen in the patchy hard
ellipse fluid. Upon further compression, both shapes undergo a stark
structural crossover. As the spatial limitations become
severe, $f(\varphi)$ for both ellipses ($n=2$) and
rectangle-like superellipses ($n=10$) systematically
transforms to exhibit sharp, dominant peaks at the vertical
orientations ($\varphi=0$ and $\pm\pi$, where the major
axis is perpendicular to the channel axis).
Concurrently, $X_u$ increases significantly within this
regime, showing that fewer particles remain bonded.
This behaviour confirms that both shapes are sterically
forced into a vertically ordered state to minimise the
occupied length along the channel axis, completely breaking
the A--B bonds.

\subsubsection{Thermodynamic Properties}
\begin{figure}[!ht]
\centering
\subfigure[]{
\includegraphics[scale=.95]{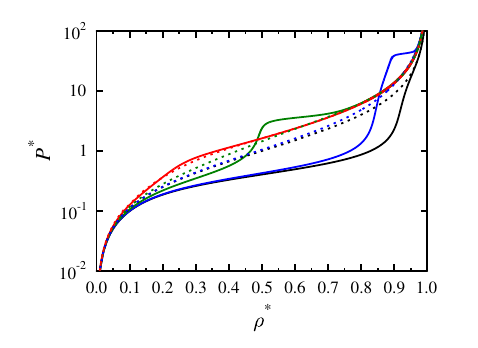}
\label{4a}
}

\subfigure[]{
\includegraphics[scale=.95]{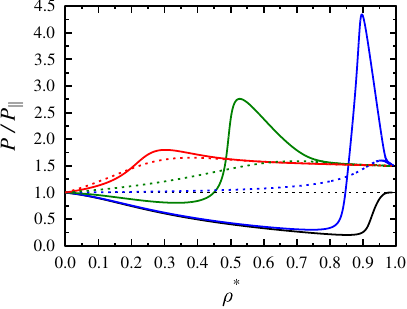}
\label{4b}
}

\caption{Equation of state of patchy hard ellipses. Reduced pressure ($P^*$) and pressure ratio ($P/P_{\parallel}$) as a function of reduced density are shown in panels (a) and (b), respectively. Curves for hard-body reference systems are shown as dashed lines, while those for patchy systems are shown as solid lines. The following aspect ratios are considered: $k = 1$ (black), $k = 1.1$ (blue), $k = 2$ (green), and $k = 5$ (red). The A--B bond is specified with $\beta \varepsilon = 5$, $\delta = 0.2$, and $\varphi_c = 20$.}
\label{fig:pressurenm2}
\end{figure}

\textit{Ellipses:}
We continue with the thermodynamic properties for patchy
hard ellipses ($n=2$) with A--B bonding.
Fig~\ref{fig:pressurenm2} illustrates the equation of
state (EOS) in two forms: a) pressure versus
density and b) pressure ratio ($P/P_{\parallel}$)
versus density, where $P_{\parallel}$ represents the pressure of parallel hard
ellipses with 2 \textit{a} lengths along the
channel axis dictated by the theoretical limit of maximum packing efficiency. In the pressure versus density representation
(Fig.~\ref{4a}), the unpatched hard-disk
reference system ($k=1$, $n=2$, $\varepsilon=0$) and the
patchy disk system ($k=1$, $n=2$, $\varepsilon\neq0$) show
a monotonic growth lacking specific non-monotonic features.
However, $k>1$ produces a
dramatic change in the curves, especially for very small elongation ($k=1.1$), where a pronounced plateau
develops at intermediate densities.
This plateau serves as a clear thermodynamic signature of
the structural rearrangement, in which the system
transforms from the tilted, bonded chain structure into the
unbonded, vertically aligned packing configuration, as
verified in Fig~\ref{fig:odf_xu}.
As $k$ is increased further ($k=2$ and $k=5$), this plateau
feature gradually weakens and the EOS curve shifts,
approaching the behaviour of the unpatched hard-ellipse
reference fluid at high pressures.
This result indicates that steric excluded-volume
interactions progressively overwhelm the energetic
contributions of the patchy attractions as the particles
become increasingly elongated. The competitive interplay between bonding, orientational
entropy, and packing entropy is more pronounced in the
$P/P_{\parallel}$ vs.\ $\rho$ plane, as shown in
Fig~\ref{4b}.
At very low densities, the ordering of patchy ellipses is
approximately isotropic due to the dominance of
orientational entropy.
As density increases, bond energy and packing entropy begin
to dominate, favouring strongly ordered configurations to
minimise the occupied length and to maximise the number of
bonds along the channel axis.
Due to the competition between bond energy, orientational
entropy, and packing entropy, the continuous structural
ordering gives rise to a distinct $P/P_{\parallel}$ peak at
intermediate densities.
The position of the peak systematically shifts toward lower
densities as $k$ increases, a trend that is
also present in the unpatched hard-ellipse fluids (dashed
lines).
This trend mirrors the behaviour of higher-dimensional
liquid-crystal-forming systems, where the
isotropic--nematic (IN) phase transition shifts to lower
densities as the particles become more elongated.
Although a true thermodynamic IN phase transition is
suppressed in our q1D system, the observed
$P/P_{\parallel}$ peak serves as a clear thermodynamic
marker of orientational structural crossovers, which can
be interpreted as a one-dimensional remnant of the
higher-dimensional IN phase transition.
Note that the patch--patch interaction makes this ordering
more complex because it acts directly against the isotropic
state at low densities; consequently, instead of a simple
IN-like structural change, the ordering emerges as a sharp
competition between the tilted bonded chain structures and
the vertically aligned ordered structures. As the close-packing limit ($\rho\rightarrow 1$) is
approached, the pressure ratio reaches different values for
$k=1$ and $k>1$.
For the $k=1$ patchy hard-disk system, the pressure ratio
approaches unity, whereas for all elongated patchy ellipses
($k>1$), it asymptotically reaches a value of $1.5$.
These close-packing values remain completely unaffected by
the presence of the A--B bonds, confirming that
the close-packing thermodynamic limit is entirely dictated
by the change in the symmetry of the particle shape.

\begin{figure}[!ht]
  \centering
  \includegraphics[scale=.95]{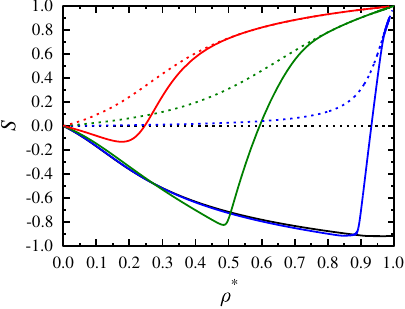}
  \caption{Orientational ordering of patchy hard ellipses. Order parameter as a function of reduced density. Curves of hard-body reference systems are shown as dashed lines, while those of patchy systems as solid lines. The following aspect ratios are considered: $k=1$ (black), $k=1.1$ (blue), $k=2$ (green) and $k=5$ (red). A--B bond is specified with $\beta \varepsilon = 5$, $\delta = 0.2$, $\varphi_c = 20$.}
  \label{fig:ordernm2}
\end{figure}

The extent of orientational ordering is measured with the order parameter, $S=\langle\cos(2\varphi)\rangle$,
which is $-1$ for perfect horizontal ordering and $+1$ for
perfect vertical ordering.
The density dependence of $S$ is shown in
Fig~\ref{fig:ordernm2}. For the unpatched hard-disk system ($k=1$, $n=2$,
$\varepsilon=0$), $S$ remains zero across all
densities because the particles possess continuous
rotational symmetry.
For the patchy hard-disk system ($k=1$, $n=2$,
$\varepsilon\neq0$), the order parameter drops to negative
values at intermediate densities, indicating a strong
preference for alignment along the channel axis
($\varphi=\pm\pi/2$), where the opposite attractive tips bind to form long bonded chains.
Crucially, the presence of the A--B bonds alters
the symmetry of the patchy disk fluid by breaking its
continuous rotational symmetry, thereby inducing a
directional, polar character.
We note that the system never reaches perfect horizontal
order ($S=-1$) because the A--B bonds possess a
finite patch opening angle ($\varphi_c=20^\circ$) that
permits local orientational fluctuations.
Under the assumption that all particles are confined
uniformly within the $2\varphi_c$ angular interval along
the channel axis, one can derive an approximate analytical
close-packing limit of
$S = -\sin(2\varphi_c)/(2\varphi_c) \approx -0.921$.
This value is in very good agreement with the exact
numerical TOM results, which yield a saturation value of
$S\approx -0.9205$.
The small difference arises because the analytical formula
assumes that all particles are bonded, whereas the exact
results show that a non-zero fraction of particles lie
outside the $2\varphi_c$ angular interval and do not
participate in bonds. For elongated particles ($k>1$),
at low-to-intermediate densities, the patch attractions remain
strong enough to stabilise the bonded networks along the
channel, forcing $S$ into the negative regime and
establishing a pronounced near-horizontal alignment.
At these densities, the particles prefer to organise into tilted
structures characterised by orientations
$\pm\pi/2\pm\varphi_c$.
Because these tilted configurations remain relatively close
to the horizontal orientation, their effects are not
visible in the order parameter curves shown in
Fig~\ref{fig:ordernm2}. Upon further compression, packing constraints take over
entirely and $S\rightarrow 1$.
The change in $S$ from negative to positive values therefore
confirms the structural crossover from tilted to vertical
structures predicted by
Figs~\ref{fig:odf_xu} and~\ref{fig:pressurenm2}.
In stark contrast to the disk case, perfect vertical order
($S=1$) is reached at maximum close packing
($\rho^{*}\rightarrow 1$) for all $k>1$, because the
close-packing structure is entirely non-degenerate in
angle, as particles are forced to align to
$\varphi=0$ or $\pm\pi$.
Since the reference hard ellipse fluid exhibits purely
positive ordering as $S$ increases smoothly from 0, the low-density
horizontal ordering $(S<0)$ is due to
A--B bonding, whereas the high-density vertical state is packing entropy driven. Regarding the effect of increasing $k$, steric excluded-volume interactions progressively overwhelm the energetic contributions of the patchy attractions, because the minimum of $S$ becomes less negative at intermediate densities, signalling a clear weakening of the patch-stabilised bonded structures.

\begin{figure}[!ht]
  \centering
  \includegraphics[scale=.95]{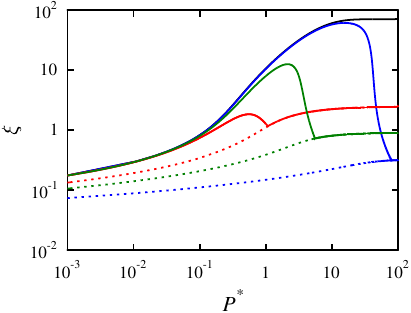}
  \caption{ Angular correlations of patchy hard ellipses. The correlation length as a function of reduced pressure. Curves of hard-body reference systems are shown as dashed lines, while those of patchy systems as solid lines. The following aspect ratios are considered: $k=1$ (black), $k=1.1$ (blue), $k=2$ (green) and $k=5$ (red). A--B bond is specified with $\beta \varepsilon = 5$, $\delta = 0.2$, $\varphi_c = 20$. The angular correlation length of hard disks $(n=2, k=1)$ is not shown since $\xi = 0$ for this system.}
  \label{fig:corrlen}
\end{figure}

The sign of the structural reorganisation is also reflected
in the behaviour of the orientational correlation length
shown in Fig~\ref{fig:corrlen}. In the unpatched hard-disk system ($k=1$, $n=2$,
$\varepsilon=0$), the correlation length is identically
zero across all densities due to the continuous rotational
symmetry.
In the patchy disk system ($k=1$, $n=2$,
$\varepsilon\neq0$), the correlation length increases
continuously with pressure and density, reaching its
maximum at close packing.
For anisotropic fluids ($k>1$, $n=2$, $\varepsilon\neq0$),
however, a peak develops at intermediate
pressures.
This peak precisely coincides with the density interval
associated with the $P/P_{\parallel}$ peak and the
order parameter crossover from negative to positive values, as the
system reorganises from horizontally bonded, tilted chains
into vertically aligned configurations. Once vertical alignment is achieved at high pressures, the
correlation length drops abruptly to a finite value
matching the hard body reference curves (dashed lines in
Fig~\ref{fig:corrlen}), indicating that orientational
correlations saturate at high pressures for ellipses.

\begin{figure}[!ht]
\centering
\subfigure[]{
\includegraphics[scale=.95]{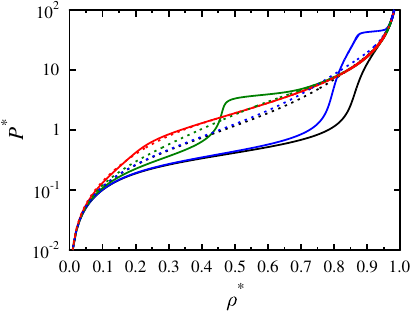}
\label{7a}
}

\subfigure[]{
\includegraphics[scale=.95]{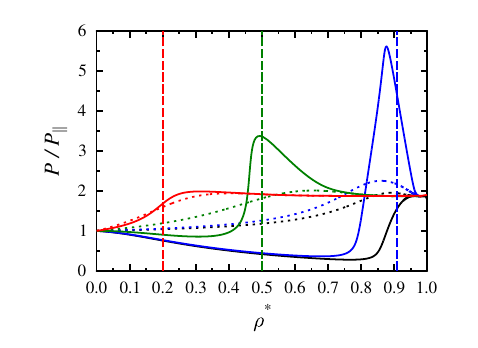}
\label{7b}
}

\caption{ Equation of state of patchy hard superellipses. Reduced pressure ($P^*$) and pressure ratio ($P/P_{\parallel}$) as a function of reduced density are shown in panels (a) and (b), respectively. Results are presented for $n=10$. Curves of hard-body reference systems are shown as dashed lines, while those of patchy systems as solid lines. The following aspect ratios are considered: $k=1$ (black), $k=1.1$ (blue), $k=2$ (green) and $k=5$ (red). A--B bond is specified with $\beta \varepsilon = 5$, $\delta = 0.2$, $\varphi_c = 20$. The vertical dashed lines denote the close packing density of horizontal ordering. }
  \label{fig:fourplots5}
\end{figure}

\begin{figure}[!ht]
  \centering
  \includegraphics[scale=.95]{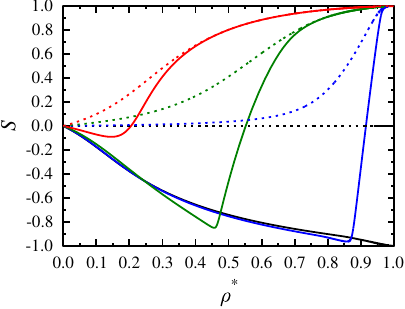}
  \caption{Orientational ordering of patchy hard superellipses. Order parameter as a function of reduced density. Curves of hard-body reference systems are shown as dashed lines, while those of patchy systems as solid lines. Results are presented for $n=10$. The following aspect ratios are considered: $k=1$ (black), $k=1.1$ (blue), $k=2$ (green) and $k=5$ (red). A--B bond is specified with $\beta \varepsilon = 5$, $\delta = 0.2$, $\varphi_c = 20$.}
 \label{fig:fourplots6}
\end{figure}

\begin{figure}[h]
  \centering
  \includegraphics[scale=.95]{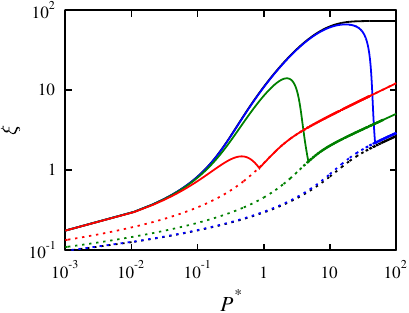}
  \caption{Angular correlations of patchy hard superellipses. The correlation length as a function of reduced pressure. Curves of hard-body reference systems are shown as dashed lines, while those of patchy systems as solid lines. Results are presented for $n=10$. The following aspect ratios are considered: $k=1$ (black), $k=1.1$ (blue), $k=2$ (green) and $k=5$ (red). A--B bond is specified with $\beta \varepsilon = 5$, $\delta = 0.2$, $\varphi_c = 20$.}
  \label{fig:fourplots7}
\end{figure}

\textit{Rectangle-like Superellipse:} Next, we study the impact of changing the particle geometry
from ellipse to rectangle-like by setting
the shape exponent to $n=10$, because the rectangle-like shape changes the rotational symmetry of the particle and the bulk properties. Note that increasing n results in straight edges and sharp corners. Figs~\ref{fig:fourplots5}, \ref{fig:fourplots6},
and~\ref{fig:fourplots7} present the corresponding exact
EOS, orientational order parameter, and correlation length
curves, respectively.

For a fluid of unpatched square-like superdisks ($k=1$,
$n=10$, $\varepsilon=0$), the continuous rotational
symmetry of a hard particle is reduced to a discrete
fourfold symmetry.
For elongated unpatched superellipses ($k>1$, $n=10$,
$\varepsilon=0$), the straight edges and sharp corners
impose a twofold symmetry.
This structural distinction is clearly visible in the
close packing limit of the pressure ratio
(Fig.~\ref{fig:fourplots5}b). As $\rho^{*}\rightarrow 1$, it is known that the vanishing
angular fluctuations generate an $n$ dependent pressure
ratio given by $P/P_{\parallel}=1-1/n$~\cite{Mizani-Oettel-Gurin-Varga_SciPostPhysCore_2025}.
For the present system ($n=10$), the pressure ratio
asymptotically approaches $P/P_{\parallel}\rightarrow 1.9$
at close packing for any value of $k$, markedly higher than the $1.5$ limit
observed for ellipses.
This indicates that thermal angular fluctuations of
rectangle-like particles cost significantly more excluded
volume along the channel axis than those of their smoothly
curved counterparts. Crucially, the presence of A--B bonds and the
varying aspect ratio do not affect the close-packing values
of the pressure ratio, meaning that the high-pressure limit
of $P/P_{\parallel}$ remains entirely determined by the
shape exponent. The effects of A--B bonding and $k$ are clearly visible
at intermediate densities with plateaus in $P$ and peaks in $P/P_{\parallel}$. In the case of hard ellipses, $P/P_{\parallel}$ peaks mark the orientational structural change from quasi-isotropic to nematic structures. When A--B bonding is also present, $P/P_{\parallel}$ peaks show the structural change from horizontal to vertical ordering, because it is quite close to the close packing density of horizontal ordering (Eq.~(\ref{eq:rhoc})) shown as a vertical dashed line in Fig.~\ref{7b}. The patch--patch interaction makes the structural
reorganisation more pronounced, which manifests as more elevated peaks in the pressure ratio. The height differences between the $n=2$ and $n=10$ cases directly demonstrate that the structural change is amplified when particles possess straight edges and sharp corners, as the straight sides introduce stronger steric packing constraints that force the system to suppress its orientational freedom more intensely and abruptly in order to satisfy the spatial limitations.

The increasing $n$ has an effect on the order parameter, too (Fig.~\ref{fig:fourplots6}).
Remarkably, for the patchy square-like superdisk fluid
($k=1$, $n=10$, $\varepsilon\neq0$), the order parameter
decreases monotonically with density and reaches
perfect horizontal order at the close packing density, where $S=-1$.
This represents a phenomenon completely inaccessible to
patchy hard disks.
In the patchy hard-disk system, particles can continuously
rotate within the finite bond opening angle without
breaking the A--B bonds, maintaining a
degenerate close packing. In contrast, the straight edges of the square-like
superdisks completely suppress the angular degeneracy at close packing.
To optimise packing with A--B bonds, the particles are locked into a
non-degenerate configuration in which every particle is forced into a perfectly horizontal orientation
($\varphi=\pm\pi/2$). For fluids of patchy superellipses ($k>1$, $n=10$,
$\varepsilon\neq0$), the minimum value of $S$ becomes less negative and $S$ converges to the unpatched hard
superellipse order parameter (dashed curve in
Fig.~\ref{fig:fourplots6}) at lower densities with
increasing $k$.
These results follow the trend observed in the patchy ellipse system, confirming that
enhanced elongation works against the energetically
stabilised horizontally bonded chains by favouring
vertical packing.
The order parameter also indicates a structural crossover
from horizontal to vertical ordering as $S$ changes from negative to positive values, and this crossover systematically weakens with increasing $k$.
As in the case of $n=2$, $S$ converges to
perfect vertical order ($S\rightarrow 1$) at close-packing density due to the non-degenerate nature of the close-packed structure.

Finally, the orientational correlation length for the
rectangle-like patchy superellipse system
(Fig.~\ref{fig:fourplots7}) reveals another major
difference from the ellipse case.
For ellipses ($n=2$), the correlation length exhibits a
finite peak at intermediate pressures before dropping and
saturating at high pressures, reflecting a local
decoupling of orientational correlations once the bonded
chains are sterically disrupted.
For rectangle-like patchy superellipses ($n=10$), a finite
peak also appears at intermediate pressures signalling the
horizontal-to-vertical structural change; however, the
correlation length diverges with pressure as
$\xi\sim P^{1/2-1/n}$ without $k$ dependence~\cite{mizani2025competition}.
For the present system ($n=10$), the correlation length
scales with pressure as $\xi\sim P^{0.4}$ in the
high-pressure vertical structure.
This power-law divergence stems from the fact that any
local angular fluctuation away from perfect vertical
alignment incurs a significantly higher excluded-distance
cost along the channel axis when particles possess straight
boundaries rather than smoothly curved shapes~\cite{mizani2025competition}.
Consequently, small orientational fluctuations cannot be
easily absorbed or damped in the vertical structure;
instead, the fluctuation propagates through the channel
over long spatial distances.
Interestingly, while this scaling law applies broadly, the
divergence of $\xi$ is not visible within the plotted
pressure range for the square-like patchy superdisk fluid
($k=1$, $n=10$, $\varepsilon\neq0$), because its
structural change occurs at much higher pressures than
those shown in Fig.~\ref{fig:fourplots7}.
For all elongated systems ($k>1$), however, the structural
transition occurs at lower pressures, changing the system
from a horizontal chain fluid with a local maximum in the
correlation length into a vertical fluid with a diverging
correlation length.
\subsection{Comparison TOM results  with PL+WTPT1}

Now we compare our exact TOM results with those of the
PL theory for hard ellipses
and the PL theory combined with generalised WTPT1 for patchy hard ellipses.
This systematic comparison allows us to determine how
accurately these theoretical approaches capture the sharp structural crossovers, equations of state, and
orientational order parameter in q1D confinement.

\begin{figure}[ht!]
\centering
\subfigure[]{
\includegraphics[scale=.95]{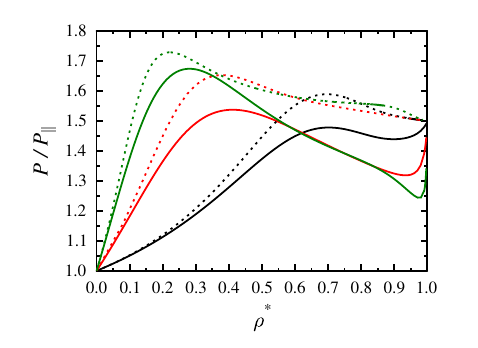}
\label{10a}
}

\subfigure[]{
\includegraphics[scale=.95]{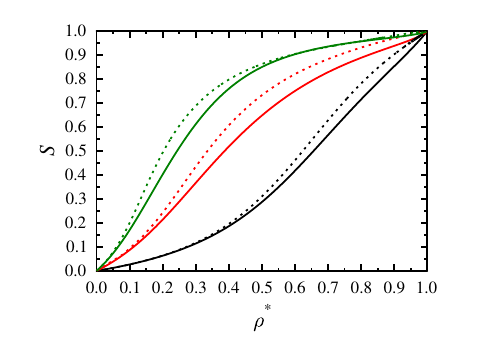}
\label{10b}
}

 \caption{Parsons--Lee theory of hard ellipses. Pressure ratio $P/P_{\parallel}$ and order parameter as a function of reduced density in panels (a) and (b), respectively. Curves of TOM results are shown as dashed lines, while those of PL theory as solid lines for $k=2$ (black), $k=5$ (red), and $k=10$ (green).}
  \label{fig:twoplots1}
\end{figure}

 Fig~\ref{fig:twoplots1} presents the density dependence of the pressure ratio
($P/P_{\parallel}$) and the order parameter
($S$) for hard ellipses ($n=2$,
$k>1$, $\varepsilon=0$). It can be seen that the PL theory is quite reliable and
captures the essential physics of the hard-ellipse
fluid when compared with the exact TOM results.
The theory correctly predicts the systematic shift of the
$P/P_{\parallel}$ peak toward lower densities as $k$ is increased, verifying its capacity to
describe the IN structural change.
Furthermore, the PL approach successfully captures the
growth of $S$, predicting the entropy-driven transition
from the quasi-isotropic low-density structure to the vertically ordered high-density one. Despite its successes, two weaknesses of the PL
theory are visible in Fig~\ref{fig:twoplots1}.
First, the PL theory slightly underestimates $S$ across the
entire density range, indicating that the packing entropy
gain arising from orientational ordering is not fully
captured. Second, the PL theory systematically
underestimates the pressure ratio $P/P_{\parallel}$ except
at the ideal-gas limit ($\rho^{*}\rightarrow 0$) and the close-packing limit ($\rho^{*}\rightarrow 1$), showing that the hard-body exclusion contribution to the excess free energy is underestimated by the PL approximation.

\begin{figure}[!ht]
\centering
\subfigure[]{
\includegraphics[scale=.95]{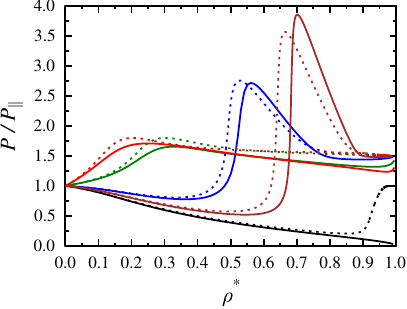}
\label{11a}
}

\subfigure[]{
\includegraphics[scale=.95]{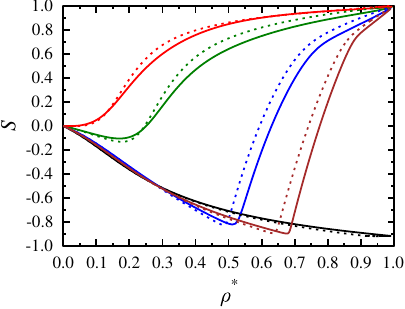}
\label{11b}
}

\subfigure[]{
\includegraphics[scale=.95]{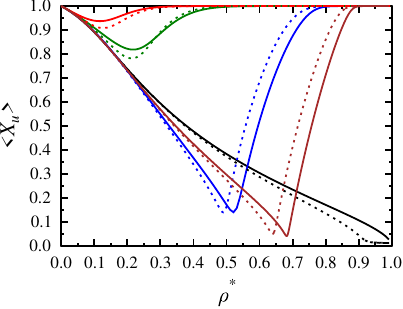}
\label{11c}
}

\caption{PL+WTPT1 for patchy hard ellipses. Pressure ratio $P/P_{\parallel}$, order parameter, and average unbonded fraction of sites are shown as functions of reduced density. Curves of TOM results are shown as dashed lines, while those of PL+WTPT1 are shown as solid lines for: $k=1$ (black), $k=1.5$ (brown), $k=2$ (blue), $k=5$ (green), and $k=10$ (red). A--B bond is specified with $\beta \varepsilon = 5$, $\delta = 0.2$, $\varphi_c = 20$.}
  \label{fig:threeplots}
\end{figure}

Fig~\ref{fig:threeplots} illustrates the accuracy of
the PL+WTPT1 theory for the A--B bonding
hard-ellipse system, showing its predictions for the
pressure ratio $P/P_{\parallel}$, the orientational order parameter $S$, and the orientationally averaged fraction of sites
not in a bond $\langle X_u \rangle$, compared against the exact TOM curves. Across the full density range, the theory provides a
qualitatively reasonable description of the structural
properties: it predicts the $P/P_{\parallel}$ peak, the
negative well and the positive saturation of $S$, and the
minimum value of $\langle X_u \rangle$, confirming that
PL+WTPT1 successfully accounts for the stabilisation of
horizontally aligned bonded chains and vertically aligned
non-bonded ellipses.
Fig~\ref{fig:threeplots} also shows that the PL+WTPT1
correctly predicts the shift of the horizontal-to-vertical
structural change to lower densities with increasing $k$,
and that the structural change weakens with increasing $k$. However, several quantitative weaknesses remain.
The PL+WTPT1 systematically overestimates the density at
which the $P/P_{\parallel}$ peak and the minima of $S$ and
$\langle X_u \rangle$ occur, shifting the predicted
structural change to slightly higher densities.
Moreover, the extent of orientational ordering is
underestimated by PL+WTPT1, and the positive rise of $S$
under compression begins at higher densities than observed
in the exact TOM results.
Consequently, the $\langle X_u \rangle$ vs.\ $\rho^{*}$
curves show that the PL+WTPT1 underestimates the number of
bonds at densities below the structural crossover
and overestimates the number of bonds above it.
In the case of patchy hard disks, $\langle X_u \rangle$
does not reach zero before the close-packing density, but
approaches 1 at $\rho^{*}\rightarrow 1$.
The pressure ratio further reveals the failure of PL+WTPT1
for patchy hard disks: the exact TOM gives
$P/P_{\parallel}\rightarrow 1$ as $\rho^{*}\rightarrow 1$,
whereas PL+WTPT1 incorrectly predicts
$P/P_{\parallel}\rightarrow 0$ in the same limit.
As this difference emerges from the patch-patch interaction, not only the PL approximation, but also WTPT1 gives rise to both qualitative and quantitative errors in the thermodynamic and structural properties of patchy hard particles.

 \section{Conclusion}

In this work, we have investigated the interplay
between directional self-assembly, geometric anisotropy,
and spatial confinement in q1D
fluid channels.
By employing the exact Transfer Operator Method (TOM) and
the approximate Parsons--Lee (PL) theory for hard-body
fluids combined with the first-order thermodynamic
perturbation theory of association (WTPT1), we have studied
the thermodynamic and orientational properties of patchy
superellipses confined in a q1D channel.

The exact TOM calculations reveal that particle shape and
directional attractions together dictate the fluid structure
across the entire density range when the particle is not
elongated ($k=1$).
However, for elongated particles ($k>1$), this statement
holds only at low and intermediate densities, since the
particle shape alone does not determine the close-packing
structure.
For patchy hard ellipses ($n=2$), the system undergoes a
structural crossover upon compression, moving from a tilted
bonded chain structure to an unbonded, vertically aligned
packed structure.
In contrast, for rectangle-like patchy superellipses
($n=10$), this behaviour manifests as a
horizontal-to-vertical structural change.
In both systems, the A--B bond actively works
against isotropic ordering at low and intermediate
densities, forcing the particles to adopt horizontal
orientations and to form chains along the channel.
At high densities, however, packing constraints break the A--B bonds in both systems to minimise the
occupied length along the channel.
As a result, both ellipses and superellipses are forced into vertical
alignment, losing the possibility of A--B bonding.
This structural change is clearly reflected macroscopically
in the equation of state: (1) a plateau emerges in the
pressure versus density curve, and (2) a peak arises in
the pressure ratio versus density curve.
In the absence of competing structures, neither plateaus
nor peaks appear in these curves for $k=1$.

The particle shape and spatial confinement also introduce
stark differences in the behaviour of the orientational
order parameter and the correlation length.
For the patchy hard-disk system ($k=1$, $n=2$), the A--B
bond does not break within the $2\varphi_c$ angular window,
meaning that perfect horizontal order is never achieved at
close packing.
Conversely, the straight edges and sharp corners of patchy
square-like superdisks ($k=1$, $n=10$) completely eliminate
this angular tilt degeneracy, locking the A--B bonding
particles into an array along the channel.
The resulting non-degenerate horizontal close-packing
structure is characterised by perfect horizontal order with
$S\rightarrow -1$.
For elongated particles ($k>1$), the close-packing
structure is vertical without A--B bonds, giving
$S\rightarrow 1$.
For patchy hard ellipses ($n=2$), the orientational
correlation length remains finite at high pressures
because the cost of angular fluctuations in the excluded
distance is weak (quadratic).
For rectangle-like patchy hard particles ($n>2$), however,
the cost of angular fluctuations is higher, causing the
correlation length to diverge as $\xi\sim P^{1/2-1/n}$ ~\cite{mizani2025competition}.
The presence of A--B bonding also increases the correlation
length, but the rearrangement of particles into vertical
order disrupts the long connected chains, resulting in a
maximum in the correlation length.
For patchy hard disks, the correlation length saturates
because the finite patch-angle window permits finite
angular fluctuations without bond disruption.

A fascinating perspective concerns the possible
manifestation of glassy behaviour in the immediate vicinity
of the maximum close-packing density.
In the fluid of patchy hard ellipses, the system can
smoothly escape the intermediate tilted configuration and
realign vertically, because the smooth curvature of the
ellipse boundary allows particles to break their selective
bonds with rotation into vertical alignment without causing steric overlaps.
This kinetic route is fundamentally blocked in systems of
patchy rectangle-like superellipses due to the rigid
constraints imposed by their straight edges and sharp
corners.
In this system, all particles within a chain must share the
same orientation, since the A--B bond renders the
symmetry polar.
The system therefore consists of two types of chains:
(1) chains in which all particles are oriented at
$\varphi=-\pi/2$ (right-oriented chains), and (2) chains
in which all particles are oriented at $\varphi=\pi/2$
(left-oriented chains).
To break these chains at very high densities, a simple
rotation of a particle toward the vertical configuration
is sterically prohibited, as the rotating corners
inevitably generate hard-body overlaps with neighbouring
particles.
Simultaneously, breaking the selective bonds via horizontal
translational motion is highly unlikely because all
available free distance along the channel effectively
vanishes in the vicinity of close packing.
Rotating a vertical particle back to the horizontal
direction is even more difficult, as it requires longer
free distance along the channel, and the rotation
direction must be $+\pi/2$ for a left-oriented neighbouring
particle or $-\pi/2$ for a right-oriented one.
This reorientation problem is exceptionally severe for
patchy hard squares ($k=1$, $n=10$), because the
close-packed structure of chain-forming patchy squares is
geometrically identical to the close-packing configuration
of unpatched hard squares, meaning that the A--B bond does
not eliminate the close-packing degeneracy.
If a defect particle freezes in a vertical orientation
within a dense horizontal chain-forming fluid, it cannot
easily rotate back to form a bond with the right- or
left-oriented neighbouring particle.
Even a successful $\pi/2$ rotation leaves the particle as
a defect if both neighbours do not share the same
horizontal orientation.
Therefore, to eliminate a defect and repair the chain, the
rotation must occur at the correct angle, which is a
formidable task owing to steric effects and random
orientational fluctuations.
This mechanism shares similarities with the structural
bottlenecks observed in fluids of hard disks confined
between two parallel walls, where the formation of a
defect-free zigzag packing structure is kinetically
hindered if neighbouring disks are locked to the same
wall~\cite{godfrey2014static}.
However, a profound topological difference exists between
these two systems.
In the confined hard-disk fluid, a defect is inherently
positional and requires two particles to constitute a
defect pair.
In the fluid of chain-forming patchy squares, the defect is
strictly orientational and resides entirely within a single
particle.
Investigating the dynamical and relaxation properties of these patchy superdisk systems via molecular dynamics simulations represents an exceptionally promising future direction to unravel the exact role of directional bonding in low-dimensional glassy states ~\cite{robinson2016glasslike,arenzon2011glassy}.

This work provides a rigorous foundation for
evaluating the reliability of perturbation theories widely
used in higher-dimensional systems.
Our comparisons demonstrate that both the PL approximation
and the combined PL+WTPT1 perturbation theory are highly
reliable in one dimension, accurately capturing the
thermodynamic properties and tracking the structural
crossovers.
It is noteworthy that the PL theory not only yields reliable results for q1D hard ellipse fluids, but also accurately describes the equations of state and IN properties of higher-dimensional (two- and three-dimensional) hard body fluids.~\cite{aliabadi2022capillary,marechal2017density,gamez2010parsons}.
The present PL+WTPT1 approach is novel in that it extends
Wertheim's first-order thermodynamic perturbation theory by
treating the fraction of sites not in a bond as an
orientation-dependent quantity, $X_u(\varphi)$, rather
than assuming $X_u$ to be uniform across all orientations.
In light of its excellent agreement with the exact TOM
results, we expect that the orientationally resolved
PL+WTPT1 theory will serve as a powerful tool for
higher-dimensional patchy systems, where exact results
cannot be obtained by statistical-mechanical methods.
A natural and compelling playground for this theory will be the system of patchy hard rods or cylinders, which undergo rich liquid-crystalline and associative phase transitions in two and three dimensions~\cite{repula2019directing}.
While the TOM is applicable only to q1D fluids, our
generalised PL+WTPT1 theory remains dimensionally
unrestricted and provides a reliable theoretical framework
for studying self-assembling materials with diverse patch
architectures.

\section*{Acknowledgements}
S. M. acknowledges the financial support of Humboldt Foundation. P. G. and S. V. gratefully acknowledge the financial support of the National Research, Development, and Innovation Office - NKFIH K137720.

\bibliography{references}

@article{arenzon2011glassy,
  title={Glassy dynamics and hysteresis in a linear system of orientable hard rods},
  author={Arenzon, Jeferson J and Dhar, Deepak and Dickman, Ronald},
  journal={Physical Review E—Statistical, Nonlinear, and Soft Matter Physics},
  volume={84},
  number={1},
  pages={011505},
  year={2011},
  publisher={APS},
   doi={https://doi.org/10.1103/PhysRevE.84.011505}  
}

@article{mizani2025emergence,
  title={Emergence of mixed orientational ordering in quasi-one-dimensional superdisk and superball fluids},
  author={Mizani, Sakineh and Oettel, Martin and Gurin, P{\'e}ter and Varga, Szabolcs},
  journal={Journal of Molecular Liquids},
  pages={129023},
  year={2025},
  publisher={Elsevier},
  doi={10.1016/j.molliq.2025.129023}  
}

@article{Parsons1979,
  author = {J. D. Parsons},
  year = {1979},
  title = {Nematic ordering in a system of rods},
  journal = {Phys. Rev. A},
  volume = {19},
  number = {3},
  pages = {1225},
  doi={https://doi.org/10.1103/PhysRevA.19.1225}  
}

@article{Lee1987,
  author = {S.-D. Lee},
  year = {1987},
  title = {A numerical investigation of nematic ordering based on a simple hard-rod model},
  journal = {J. Chem. Phys.},
  volume = {87},
  number = {8},
  pages = {4972},
  doi={https://doi.org/10.1063/1.452811} 
   

}

@article{Lee1988,
  author = {S.-D. Lee},
  year = {1988},
  title = {The Onsager-type theory for nematic ordering of finite-length hard ellipsoids},
  journal = {J. Chem. Phys.},
  volume = {89},
  number = {11},
  pages = {7036},
  doi={https://doi.org/10.1063/1.455332} 
}

@article{marechal2017density,
  title={Density functional theory and simulations of colloidal triangular prisms},
  author={Marechal, Matthieu and Dussi, Simone and Dijkstra, Marjolein},
  journal={The Journal of chemical physics},
  volume={146},
  number={12},
  year={2017},
  publisher={AIP Publishing},
doi={10.1063/1.4978502}
}

@article{aliabadi2022capillary,
  title={Capillary-driven biaxial planar and homeotropic nematization of hard cylinders},
  author={Aliabadi, Roohollah and Nasirimoghadam, Soudabe and Wensink, Henricus Herman},
  journal={Physical Review E},
  volume={105},
  number={6},
  pages={064704},
  year={2022},
  publisher={APS},
doi={10.1103/PhysRevE.105.064704}
}

@article{gamez2010parsons,
  title={Parsons--Lee approach for oblate hard spherocylinders},
  author={G{\'a}mez, F and Merkling, PJ and Lago, S},
  journal={Chemical Physics Letters},
  volume={494},
  number={1-3},
  pages={45--49},
  year={2010},
  publisher={Elsevier},
doi={10.1016/j.cplett.2010.05.094}
}

@article{godfrey2014static,
  title={Static and dynamical properties of a hard-disk fluid confined to a narrow channel},
  author={Godfrey, MJ and Moore, MA},
  journal={Physical Review E},
  volume={89},
  number={3},
  pages={032111},
  year={2014},
  publisher={APS},
doi={10.1103/PhysRevE.89.032111}
}

@article{robinson2016glasslike,
  title={Glasslike behavior of a hard-disk fluid confined to a narrow channel},
  author={Robinson, JF and Godfrey, MJ and Moore, MA},
  journal={Physical Review E},
  volume={93},
  number={3},
  pages={032101},
  year={2016},
  publisher={APS},
doi={10.1103/PhysRevE.93.032101}
}

@article{repula2019directing,
  title={Directing liquid crystalline self-organization of rodlike particles through tunable attractive single tips},
  author={Repula, Andrii and Oshima Menegon, Mariana and Wu, Cheng and van Der Schoot, Paul and Grelet, Eric},
  journal={Physical Review Letters},
  volume={122},
  number={12},
  pages={128008},
  year={2019},
  publisher={APS},
doi={10.1103/PhysRevLett.122.128008}
}

@article{mizani2025competition,
  title={Competition between shape anisotropy and deformation in the ordering and close packing properties of quasi-one-dimensional hard superellipse fluids},
  author={Mizani, Sakineh and Oettel, Martin and Gurin, P{\'e}ter and Varga, Szabolcs},
  journal={Physical Review E},
  volume={111},
  number={6},
  pages={064121},
  year={2025},
  publisher={APS},
doi={10.1103/jszf-cvdn}
}

@article{Cuesta-Sanchez_JStatPhys_2004,
  author  = {Cuesta, Jos\'{e} and S\'{a}nchez, Angel},
  title   = {General non-existence theorem for phase transitions
             in one-dimensional systems with short range
             interactions, and physical examples of such
             transitions},
  journal = {Journal of Statistical Physics},
  volume  = {115},
  pages   = {869--893},
  year    = {2004},
  doi     = {10.1023/B:JOSS.0000022373.63640.4e}
}

@article{ Mizani-Oettel-Gurin-Varga_SciPostPhysCore_2025,
  author  = {Mizani, Sakineh and Oettel, Martin and Gurin,
             P\'{e}ter and Varga, Szabolcs},
  title   = {Universality of the close packing properties and
             markers of isotropic-to-tetratic crossover in
             quasi-one-dimensional superdisk fluid},
  journal = {SciPost Physics Core},
  volume  = {8},
  pages   = {008},
  year    = {2025},
  doi     = {10.21468/SciPostPhysCore.8.1.008}
}

@article{Glotzer2007,
  author  = {Glotzer, Sharon  and Solomon, Michael },
  title   = {Anisotropy of building blocks and their assembly
             into complex structures},
  journal = {Nature Materials},
  volume  = {6},
  pages   = {557--562},
  year    = {2007},
  doi     = {10.1038/nmat1949}
}

@article{Zhang2014,
  author  = {Zhang, Shuang-Yuan and Regulacio, Michelle 
             and Han, Ming-Yong},
  title   = {Self-assembly of colloidal one-dimensional nanocrystals},
  journal = {Chemical Society Reviews},
  volume  = {43},
  pages   = {2301--2323},
  year    = {2014},
  doi     = {10.1039/C3CS60397K}
}

@article{Sacanna2013,
  author  = {Sacanna, Stefano and Pine, David  and Yi, Gi-Ra},
  title   = {Engineering shape: {The} novel geometries of
             colloidal self-assembly},
  journal = {Soft Matter},
  volume  = {9},
  pages   = {8096--8106},
  year    = {2013},
  doi     = {10.1039/C3SM50500F}
}

@article{Bianchi2011,
  author  = {Bianchi, Emanuela and Blaak, Ronald
             and Likos, Christos },
  title   = {Patchy colloids: state of the art and perspectives},
  journal = {Physical Chemistry Chemical Physics},
  volume  = {13},
  pages   = {6397--6410},
  year    = {2011},
  doi     = {10.1039/C0CP02296A}
}

@article{DeMichele2012,
  author  = {De Michele, Cristiano and Bellini, Tommaso
             and Sciortino, Francesco},
  title   = {Self-assembly of bifunctional patchy particles with anisotropic shape into polymers chains: Theory, simulations, and experiments},
  journal = {Macromolecules},
  volume  = {45},
  pages   = {1090--1106},
  year    = {2012},
  doi     = {10.1021/ma201962x}
}

@article{Lu2004,
  author  = {L\"{u}, Xiaoying and Kindt, James },
  title   = {{Monte Carlo} simulation of the self-assembly and
             phase behavior of semiflexible equilibrium polymers},
  journal = {The Journal of Chemical Physics},
  volume  = {120},
  pages   = {10328--10338},
  year    = {2004},
  doi     = {10.1063/1.1729855}
}

@article{Lu2006,
  author  = {L\"{u}, Xiaoying and Kindt, James },
  title   = {Theoretical analysis of polydispersity in the nematic
             phase of self-assembled semiflexible chains},
  journal = {The Journal of Chemical Physics},
  volume  = {125},
  pages   = {054909},
  year    = {2006},
  doi     = {10.1063/1.2234478}
}

@article{Bianchi2006,
  author  = {Bianchi, Emanuela and Largo, Julio and Tartaglia, Piero
             and Zaccarelli, Emanuela and Sciortino, Francesco},
  title   = {Phase diagram of patchy colloids: {Toward} empty liquids},
  journal = {Physical Review Letters},
  volume  = {97},
  pages   = {168301},
  year    = {2006},
  doi     = {10.1103/PhysRevLett.97.168301}
}

@article{Avvisati2015,
  author  = {Avvisati, Guido and Vissers, Teun
             and Dijkstra, Marjolein},
  title   = {Self-assembly of patchy colloidal dumbbells},
  journal = {The Journal of Chemical Physics},
  volume  = {142},
  pages   = {084905},
  year    = {2015},
  doi     = {10.1063/1.4913369}
}

@article{Nguyen2018,
  title={Nematic liquid crystals of bifunctional patchy spheres},
  author={Nguyen, Khanh Thuy and De~Michele, Cristiano},
  journal={The European Physical Journal E},
  volume={41},
  number={12},
  pages={141},
  year={2018},
  publisher={Springer},
   doi    = {10.1140/epje/i2018-11750-4}
}

@article{Russo2022,
  author  = {Russo, John and Leoni, Fabio and Martelli, Fausto
             and Sciortino, Francesco},
  title   = {The physics of empty liquids: {From} patchy
             particles to water},
  journal = {Reports on Progress in Physics},
  volume  = {85},
  pages   = {016601},
  year    = {2022},
  doi     = {10.1088/1361-6633/ac42d9}
}

@article{Wang2012,
  author  = {Wang, Yufeng and Wang, Yu and Breed, Dana 
             and Manoharan, Vinothan  and Feng, Lang
             and Hollingsworth, Andrew  and Weck, Marcus
             and Pine, David },
  title   = {Colloids with valence and specific directional bonding},
  journal = {Nature},
  volume  = {491},
  pages   = {51--55},
  year    = {2012},
  doi     = {10.1038/nature11564}
}

@article{Hueckel2021,
  author  = {Hueckel, Theodore and Hocky, Glen 
             and Sacanna, Stefano},
  title   = {Total synthesis of colloidal matter},
  journal = {Nature Reviews Materials},
  volume  = {6},
  pages   = {1053--1069},
  year    = {2021},
  doi     = {10.1038/s41578-021-00323-x}
}

@article{Dijkstra2021,
  author  = {Dijkstra, Marjolein and Luijten, Erik},
  title   = {From predictive modelling to machine learning and
             reverse engineering of colloidal self-assembly},
  journal = {Nature Materials},
  volume  = {20},
  pages   = {762--773},
  year    = {2021},
  doi     = {10.1038/s41563-021-01014-2}
}

@article{Kraft2012,
  author  = {Kraft, Daniela  and Ni, Ran and Smallenburg, Frank
             and Hermes, Michiel and Yoon, Kisun and Weitz, David 
             and van Blaaderen, Alfons and Groenewold, Jan
             and Dijkstra, Marjolein and Kegel, Willem },
  title   = {Surface roughness directed self-assembly of patchy
             particles into colloidal micelles},
  journal = {Proceedings of the National Academy of Sciences},
  volume  = {109},
  pages   = {10787--10792},
  year    = {2012},
  doi     = {10.1073/pnas.1116820109}
}

@article{Morphew2018,
  author  = {Morphew, Daniel and Shaw, James and Avins, Christopher
             and Chakrabarti, Dwaipayan},
  title   = {Programming hierarchical self-assembly of patchy
             particles into colloidal crystals via colloidal molecules},
  journal = {ACS Nano},
  volume  = {12},
  pages   = {2355--2364},
  year    = {2018},
  doi     = {10.1021/acsnano.7b07633}
}

@article{Li2020,
  author  = {Li, Weiya and Palis, Herv\'{e} and M\'{e}rindol,
             R\'{e}mi and Majimel, J\'{e}r\^{o}me
             and Ravaine, Serge and Duguet, Etienne},
  title   = {Colloidal molecules and patchy particles:
             complementary concepts, synthesis and self-assembly},
  journal = {Chemical Society Reviews},
  volume  = {49},
  pages   = {1955--1976},
  year    = {2020},
  doi     = {10.1039/C9CS00804G}
}

@article{Teixeira2017,
  author  = {Teixeira, Paulo Ivo Cortez
             and Tavares, Jos\'{e} Maria},
  title   = {Phase behaviour of pure and mixed patchy colloids
             --- theory and simulation},
  journal = {Current Opinion in Colloid \& Interface Science},
  volume  = {30},
  pages   = {16--24},
  year    = {2017},
  doi     = {10.1016/j.cocis.2017.03.011}
}

@article{Sciortino2017,
  author  = {Sciortino, Francesco and Zaccarelli, Emanuela},
  title   = {Equilibrium gels of limited valence colloids},
  journal = {Current Opinion in Colloid \& Interface Science},
  volume  = {30},
  pages   = {90--96},
  year    = {2017},
  doi     = {10.1016/j.cocis.2017.06.001}
}

@article{Ruzicka2011,
  author  = {Ruzicka, Barbara and Zaccarelli, Emanuela
             and Zulian, Laura and Angelini, Roberta
             and Sztucki, Michael and Mouss\"{a}d, Abdellatif
             and Narayanan, Theyencheri and Sciortino, Francesco},
  title   = {Observation of empty liquids and equilibrium gels
             in a colloidal clay},
  journal = {Nature Materials},
  volume  = {10},
  pages   = {56--60},
  year    = {2011},
  doi     = {10.1038/nmat2921}
}

@article{Doppelbauer2010,
  author  = {Doppelbauer, G\"{u}nther and Bianchi, Emanuela
             and Kahl, Gerhard},
  title   = {Self-assembly scenarios of patchy colloidal
             particles in two dimensions},
  journal = {Journal of Physics: Condensed Matter},
  volume  = {22},
  pages   = {104105},
  year    = {2010},
  doi     = {10.1088/0953-8984/22/10/104105}
}

@article{Sato2021,
  author  = {Sato, Masahide},
  title   = {Effect of the interaction length on clusters
             formed by spherical one-patch particles on flat
             planes},
  journal = {Langmuir},
  volume  = {37},
  pages   = {4213--4221},
  year    = {2021},
  doi     = {10.1021/acs.langmuir.1c00102}
}

@article{Jonas2022,
  author  = {Jonas, Hannah and Schall, Peter and Bolhuis, Peter},
  title   = {Extended {Wertheim} theory predicts the anomalous
             chain length distribution of divalent patchy
             particles under extreme confinement},
  journal = {The Journal of Chemical Physics},
  volume  = {157},
  pages   = {094903},
  year    = {2022},
  doi     = {10.1063/5.0098882}
}

@article{Gharibi2024,
  author  = {Gharibi, Ali and Eslami, Hossein
             and M\"{u}ller-Plathe, Florian},
  title   = {Self-assembly of model three- and four-patch
             colloidal particles in two dimensions},
  journal = {Journal of Chemical Theory and Computation},
  volume  = {20},
  pages   = {6858--6869},
  year    = {2024},
  doi     = {10.1021/acs.jctc.4c00540}
}

@article{Lebowitz1983,
  author  = {Lebowitz, Joel  and Percus, Jerome },
  title   = {One-dimensional models of anisotropic fluids},
  journal = {Annals of the New York Academy of Sciences},
  volume  = {410},
  pages   = {351--359},
  year    = {1983},
  doi     = {10.1111/j.1749-6632.1983.tb23333.x}
}

@article{Marshall2015,
  author  = {Marshall, Bennett },
  title   = {Thermodynamic perturbation theory for associating
             fluids confined in a one-dimensional pore},
  journal = {The Journal of Chemical Physics},
  volume  = {142},
  pages   = {234906},
  year    = {2015},
  doi     = {10.1063/1.4922547}
}

@article{Gurin2022,
  author  = {Gurin, P\'{e}ter and Varga, Szabolcs},
  title   = {Anomalous phase behavior of quasi-one-dimensional
             attractive hard rods},
  journal = {Physical Review E},
  volume  = {106},
  pages   = {044606},
  year    = {2022},
  doi     = {10.1103/PhysRevE.106.044606}
}

@article{Herzfeld1934,
  author  = {Herzfeld, Karl  and Goeppert-Mayer, Maria},
  title   = {On the states of aggregation},
  journal = {The Journal of Chemical Physics},
  volume  = {2},
  pages   = {38--45},
  year    = {1934},
  doi     = {10.1063/1.1749355
}
}

@article{Montero2019,
  author  = {Montero, Ana  and Santos, Andr\'{e}s},
  title   = {Triangle-well and ramp interactions in
             one-dimensional fluids: {A} fully analytic exact
             solution},
  journal = {Journal of Statistical Physics},
  volume  = {175},
  pages   = {269--288},
  year    = {2019},
  doi     = {10.1007/s10955-019-02255-x}
}

@article{Fantoni2021,
  author  = {Fantoni, Riccardo and Maestre, Miguel 
             and Santos, Andr\'{e}s},
  title   = {Finite-size effects and thermodynamic limit in
             one-dimensional {Janus} fluids},
  journal = {Journal of Statistical Mechanics: Theory and
             Experiment},
  volume  = {2021},
  pages   = {103210},
  year    = {2021},
  doi     = {10.1088/1742-5468/ac2897}
}

@article{Gurin2025,
  author  = {Gurin, P\'{e}ter and Varga, Szabolcs},
  title   = {Ordering and association of patchy particles in
             a quasi-one-dimensional channel},
  journal = {Physical Review Research},
  volume  = {7},
  pages   = {033094},
  year    = {2025},
  doi     = {10.1103/1xp1-mnqx}
}

@article{Vo2003,
  author  = {V\~{o}, Th\'ai Tuy\'en
 and Chen, Lee-Jen and Robert, Marc},
  title   = {Short-range order in linear systems},
  journal = {The Journal of Chemical Physics},
  volume  = {119},
  pages   = {5607--5613},
  year    = {2003},
  doi     = {10.1063/1.1599272}
}

@article{Fantoni2017,
  author  = {Fantoni, Riccardo and Santos, Andr\'{e}s},
  title   = {One-dimensional fluids with second nearest
             neighbor interactions},
  journal = {Journal of Statistical Physics},
  volume  = {169},
  pages   = {1171--1201},
  year    = {2017},
  doi     = {10.1007/s10955-017-1908-6}
}

@article{Jackson1988,
  author  = {Jackson, George and Chapman, Walter 
             and Gubbins, Keith },
  title   = {Phase equilibria of associating fluids: {Spherical}
             molecules with multiple bonding sites},
  journal = {Molecular Physics},
  volume  = {65},
  pages   = {1--31},
  year    = {1988},
  doi     = {
    10.1080/00268978800101601}
}

@article{McGrother1997,
  author  = {McGrother, Simon and Sear, Richard and Jackson, George},
  title   = {The liquid crystalline phase behavior of
             dimerizing hard spherocylinders},
  journal = {The Journal of Chemical Physics},
  volume  = {106},
  pages   = {7315--7330},
  year    = {1997},
  doi     = {10.1063/1.473693}
}

@article{Chen2011,
  author  = {Chen, Qian and Bae, Sung Chul and Granick, Steve},
  title   = {Directed self-assembly of a colloidal kagome lattice},
  journal = {Nature},
  volume  = {469},
  pages   = {381--384},
  year    = {2011},
  doi     = {10.1038/nature09713}
}

@article{Fernandez2015,
  author  = {Fern\'{a}ndez, Marta  and Misko, Vyacheslav 
             and Peeters, Fran\c{c}ois },
  title   = {Self-assembly of {Janus} particles into helices
             with tunable pitch},
  journal = {Physical Review E},
  volume  = {92},
  pages   = {042309},
  year    = {2015},
  doi     = {10.1103/PhysRevE.92.042309}
}

@article{Wertheim1984a,
  author  = {Wertheim, Michael},
  title   = {Fluids with highly directional attractive forces.
             {I}. {Statistical} thermodynamics},
  journal = {Journal of Statistical Physics},
  volume  = {35},
  pages   = {19--34},
  year    = {1984},
  doi     = {10.1007/BF01017362}
}

@article{Wertheim1984b,
  author  = {Wertheim, Michael},
  title   = {Fluids with highly directional attractive forces.
             {II}. {Thermodynamic} perturbation theory and
             integral equations},
  journal = {Journal of Statistical Physics},
  volume  = {35},
  pages   = {35--47},
  year    = {1984},
  doi     = {10.1007/BF01017363}
}

@article{Wertheim1986a,
  author  = {Wertheim, Michael},
  title   = {Fluids with highly directional attractive forces.
             {III}. {Multiple} attraction sites},
  journal = {Journal of Statistical Physics},
  volume  = {42},
  pages   = {459--476},
  year    = {1986},
  doi     = {10.1007/BF01127721}
}

@article{Wertheim1986b,
  author  = {Wertheim, Michael},
  title   = {Fluids with highly directional attractive forces.
             {IV}. {Equilibrium} polymerization},
  journal = {Journal of Statistical Physics},
  volume  = {42},
  pages   = {477--492},
  year    = {1986},
  doi     = {10.1007/BF01127722}
}

@article{Wertheim1987,
  author  = {Wertheim, Michael},
  title   = {Thermodynamic perturbation theory of
             polymerization},
  journal = {The Journal of Chemical Physics},
  volume  = {87},
  pages   = {7323--7331},
  year    = {1987},
  doi     = {10.1063/1.453326}
}

@article{GilVillegas1997,
  author  = {Gil-Villegas, Alejandro and Galindo, Amparo
             and Whitehead, Paul  and Mills, Stuart 
             and Jackson, George and Burgess, Andrew },
  title   = {Statistical associating fluid theory for chain
             molecules with attractive potentials of variable
             range},
  journal = {The Journal of Chemical Physics},
  volume  = {106},
  pages   = {4168--4186},
  year    = {1997},
  doi     = {10.1063/1.473101}
}

@article{Muller2001,
  author  = {M\"{u}ller, Erich  and Gubbins, Keith },
  title   = {Molecular-based equations of state for associating
             fluids: {A} review of {SAFT} and related
             approaches},
  journal = {Industrial \& Engineering Chemistry Research},
  volume  = {40},
  pages   = {2193--2211},
  year    = {2001},
  doi     = {10.1021/ie000773w}
}

@article{Economou2002,
  author  = {Economou, Ioannis },
  title   = {Statistical associating fluid theory: {A}
             successful model for the calculation of
             thermodynamic and phase equilibrium properties
             of complex fluid mixtures},
  journal = {Industrial \& Engineering Chemistry Research},
  volume  = {41},
  pages   = {953--962},
  year    = {2002},
  doi     = {10.1021/ie0102201}
}

@article{Lira2022,
  author  = {Lira, Carl  and Elliott,  Richard
             and Gupta, Sumnesh and Chapman, Walter },
  title   = {{Wertheim}'s association theory for phase
             equilibrium modeling in chemical engineering
             practice},
  journal = {Industrial \& Engineering Chemistry Research},
  volume  = {61},
  pages   = {15678--15713},
  year    = {2022},
  doi     = {10.1021/acs.iecr.2c02058}
}

@article{Alsaifi2025,
  author  = {Alsaifi, Nayef },
  title   = {First-order perturbation theory for associating
             fluids in the isothermal--isobaric ensemble},
  journal = {Industrial \& Engineering Chemistry Research},
  volume  = {64},
  pages   = {21839--21849},
  year    = {2025},
  doi     = {10.1021/acs.iecr.5c01288}
}

@article{Kalyuzhnyi2026,
  author  = {Kalyuzhnyi, Yurij },
  title   = {Phase behavior and percolation of a primitive
             model of {Laponite} suspension: {Wertheim}'s
             thermodynamic perturbation theory with anisotropic
             reference particles},
  journal = {ACS Omega},
  year    = {2026},
  doi     = {10.1021/acsomega.5c12221}
}

@article{Gurin2026,
  author  = {Gurin, P\'{e}ter and Sarkadi, Zs\'{o}fia
             and Paricaud, Patrice and Varga, Szabolcs},
  title   = {Generalized {Wertheim} {TPT1} for patchy hard
             spheres confined in a one-dimensional channel},
  journal = {The Journal of Chemical Physics},
  volume  = {164},
  pages   = {124502},
  year    = {2026},
  doi     = {10.1063/5.0323803}
}

\end{document}